\documentclass[pre,showpacs,longbibliography,twocolumn]{revtex4-1}
\usepackage{graphicx}% Include figure files
\usepackage{color}% allows text in color with \textcolor{color}{text}
\usepackage{amsmath}
\usepackage{subfigure}
\usepackage{appendix}
\usepackage{soul}
\usepackage{bbm}
\usepackage[colorlinks,linkcolor=blue,citecolor=blue,urlcolor=blue,hyperindex,breaklinks]{hyperref}
\begin{document}
\title{Magnetically controlled quantum thermal devices via three nearest-neighbor coupled spin-1/2 systems}
\author{Yi-jia Yang, Yu-qiang Liu, Zheng Liu, and Chang-shui Yu}
\email{Electronic address: ycs@dlut.edu.cn}
\affiliation{School of Physics, Dalian University of Technology, Dalian 116024, China}
\date{\today}

\begin{abstract}
A quantum thermal device based on three nearest-neighbor coupled spin-1/2 systems controlled by the magnetic field is proposed. 
We systematically study the steady-state thermal behaviors of the system. When the two terminals of our system are in contact with two thermal reservoirs, respectively, the system behaves as a perfect thermal modulator that can manipulate heat current from zero to specific values by adjusting magnetic field direction over different parameter ranges, since the longitudinal magnetic field can completely block the heat transport. Significantly, the modulator can also be achieved when a third thermal reservoir perturbs the middle spin. We also find that the transverse field can induce the system to separate into two subspaces in which neither steady-state heat current vanishes, thus providing an extra level of control over the heat current through the manipulation of the initial state. 
In addition, the performance of this device as a transistor can be enhanced by controlling the magnetic field, achieving versatile amplification behaviors, in particular substantial amplification factors.
\end{abstract}

\maketitle

\section{INTRODUCTION} 
Quantum thermodynamics \cite{landsberg1956foundations,vinjanampathy2016quantum,RevModPhys.84.1045,10.1080/00018732.2018.1519981,millen2016perspective,parrondo2015thermodynamics,Hofer_2017,Mukherjee_2021,Arrachea_2023} is a cross discipline that describes the thermodynamic behavior of quantum systems. It allows us to test thermodynamic laws at the quantum level and to design quantum thermal devices that control heat current as the electric current in electricity to realize amplification, rectification, modulation, and so on \cite{RevModPhys.84.1045,10.1080/00018732.2018.1519981}. 
Recently, more and more attention has been attracted to various quantum thermal devices, such as quantum refrigerators \cite{PhysRevLett.110.256801,PhysRevE.95.012146,PhysRevE.96.012122,PhysRevE.101.012109,Nimmrichter2017quantumclassical,PhysRevE.90.052142}, quantum thermal transistors \cite{JoulainEzzahriOrdonezMiranda+2017+163+170,PhysRevLett.116.200601,10.1063/1.4979977,PhysRevA.97.052112,PhysRevA.103.052613,PhysRevE.99.032112,PhysRevB.101.245402,PhysRevApplied.16.034026,PhysRevE.106.024110,PhysRevB.101.245402,PhysRevE.98.022118,PhysRevB.101.184510,PhysRevApplied.6.054003,PRODHOMME201852,10.1063/1.4991516,PhysRevB.100.045418,PhysRevResearch.2.033285,castelli2023three}, quantum thermal diode \cite{JoulainEzzahriOrdonezMiranda+2017+163+170,PhysRevE.89.062109,PhysRevE.95.022128,PhysRevE.99.042121,PhysRevE.99.032116,PhysRevE.104.054137,PhysRevB.103.155434,Diaz_2021,PhysRevE.106.034116,PhysRevE.107.064125}, quantum batteries \cite{PhysRevLett.118.150601,PhysRevResearch.2.023095}, quantum modulators \cite{Karimi_2017,ronzani2018tunable,PhysRevE.99.062123,Yang_2022,GUO2022115275}. 

Many physical systems such as superconducting circuits \cite{maillet2020electric,PhysRevE.96.062120,PhysRevE.102.030102,PhysRevB.101.184510}, quantum dots \cite{Scheibner_2008,PhysRevB.93.161410,GUO2022115275,PhysRevLett.110.256801,PhysRevB.100.045418,10.1063/1.4979977}, cavity QED systems \cite{PhysRevB.99.035129,PhysRevE.79.041113,yu2019quantum,Yan:23}, and optomechanical setups \cite{gelbwaser2015work,seif2018thermal,PhysRevLett.112.150602,PhysRevE.80.061129,PhysRevE.94.022141,WU2021104996} are good platforms for implementing quantum thermal devices.
The coupled spin-1/2 system has been widely employed to design thermodynamic devices such as the smallest self-contained quantum refrigerator \cite{PhysRevLett.110.256801}, the smallest quantum thermal machine \cite{PhysRevE.96.012122}, the quantum heat manager \cite{PhysRevResearch.2.033285,PhysRevE.99.062123} and the thermal transistor \cite{PhysRevLett.116.200601, PhysRevE.98.022118, PhysRevB.101.245402, PhysRevA.103.052613, PhysRevE.106.024110}. 
What's more, rapid development has been made in the simulation and experiment of magnetic models,  coupled-spin models such as the Ising model \cite{PhysRevB.68.214406,NETO20131,nature09994,YUAN2021126279,PhysRevB.68.214406,PhysRevB.85.134412,PhysRevA.103.043312,PhysRevA.106.053308} and Heisenberg XXZ model \cite{e22111311,PhysRevE.99.032136,PhysRevLett.106.217206,PhysRevE.105.024120}, including their simulated systems, have a good experimental foundation \cite{2010Quantum, PhysRevB.66.075128,nature09994,2016Real,ncs41467-022-35301-6}. Trapped ions are widely used in the spin chain system \cite{2010Quantum,WU2021104996}. In order to more conveniently control the spin-spin interaction, Rydberg-dressed atom array is a friendly candidate. Rydberg-dressed technology can be used to design the long-range interaction in spin chains or lattices \cite{PhysRevX.7.041063,PhysRevLett.124.063601}, and a distance-selective Ising interaction can be achieved by coupling an off-resonant laser with the Rydberg atomic pairs \cite{PhysRevLett.128.113602}, indicating a potential and experimentally reliable application prospect in quantum thermodynamic devices.

Furthermore, the thermodynamic device can manipulate heat currents in various ways. The heat current for the thermal transistor is usually controlled by a weak heat current \cite{PhysRevLett.116.200601, PhysRevE.99.032112}. Recently, it has been shown that the decoupled systems can control steady-state heat current by preparing initial states on purpose \cite{GUO2022115275, Yang_2022}. The dark state of a system can also be used to control heat currents for quantum thermal diodes \cite{PhysRevE.106.034116}. The control of thermal current by external manipulation is attracting increasing interest. 
Sinusoidal modulation or $\pi$-flip modulation is used to realize a periodically driven thermal transistor \cite{PhysRevE.106.024110}. In a coupled superconducting qubit system, the heat current can be controlled by changing the magnetic flux \cite{Karimi_2017}, and the thermal switch is designed based on coherently driving a single qubit in superconducting circuits \cite{PhysRevE.102.030102}. Moreover, by adjusting the Rabi frequency in the system consisting of two coupled qubits coupled to two $1$D waveguides, all-optical control of the heat current can be performed in the quantum electrodynamics \cite{Yan:23}. It is also possible to realize the quantum thermal transistor with high amplification by controlling the Rabi frequency in a weak driving system \cite{PhysRevB.101.245402}.

In this paper, we design a quantum thermal device utilizing a system composed of three nearest-neighbor coupled spins immersed in an adjustable magnetic field. Each spin is also individually connected to its respective heat reservoir, as shown visually in Fig. \ref{model}. The distinct advantage of our design is the ability to manipulate the heat current by altering the orientation of the applied magnetic field. 
To understand the behavior of our system, we have employed analytical techniques, in particular the Born-Markov-secular (BMS) master equation, to derive the steady-state properties. 
One noteworthy observation is that when the middle spin is isolated from the thermal reservoir, the longitudinal field (LF), aligned with the direction of the coupling between the spins, can effectively block the heat transfer between the system and its corresponding thermal environment. By manipulating the direction of the magnetic field, we can induce an expected heat current, essentially demonstrating the capability of our system as an ideal quantum thermal modulator, where the heat current is controlled by the magnetic field orientation. 
Moreover, we have shown that the thermal modulator still works effectively even when the middle spin is connected to a perturbation reservoir. In particular, by appropriately adjusting the parameters, such as the magnetic field and coupling strength, we can completely block the heat current, highlighting the characteristic of our modulator. In the case of the transverse field (TF), where the magnetic field direction is orthogonal to the spin coupling direction, our system decouples into two independent subspaces. This observation provides an alternative approach to modulating the heat current by controlling the proportions of the initial state in each subspace.
Finally, when our system serves as a thermal transistor, we have found that the controllable magnetic field significantly enhances the transistor's performance compared to the LF scenario \cite{PhysRevLett.116.200601}. 

This paper is structured as follows. Sec. \ref{section2} gives the primary model and deals with the dynamics of the system. In Sec. \ref{section3}, we present the steady-state heat current of the system. 
In Sec. \ref{section4}, the magnetically controlled thermal devices are implemented and analyzed, specifically, the quantum thermal modulator and the quantum thermal transistor are described in Sec. \ref{section4A} and Sec. \ref{section4B}, respectively. Sec. \ref{section5} gives the discussion and conclusion.

\section{Model and steady state}
\label{section2}
\begin{figure}
	\centering
		\includegraphics[width=8.3cm]{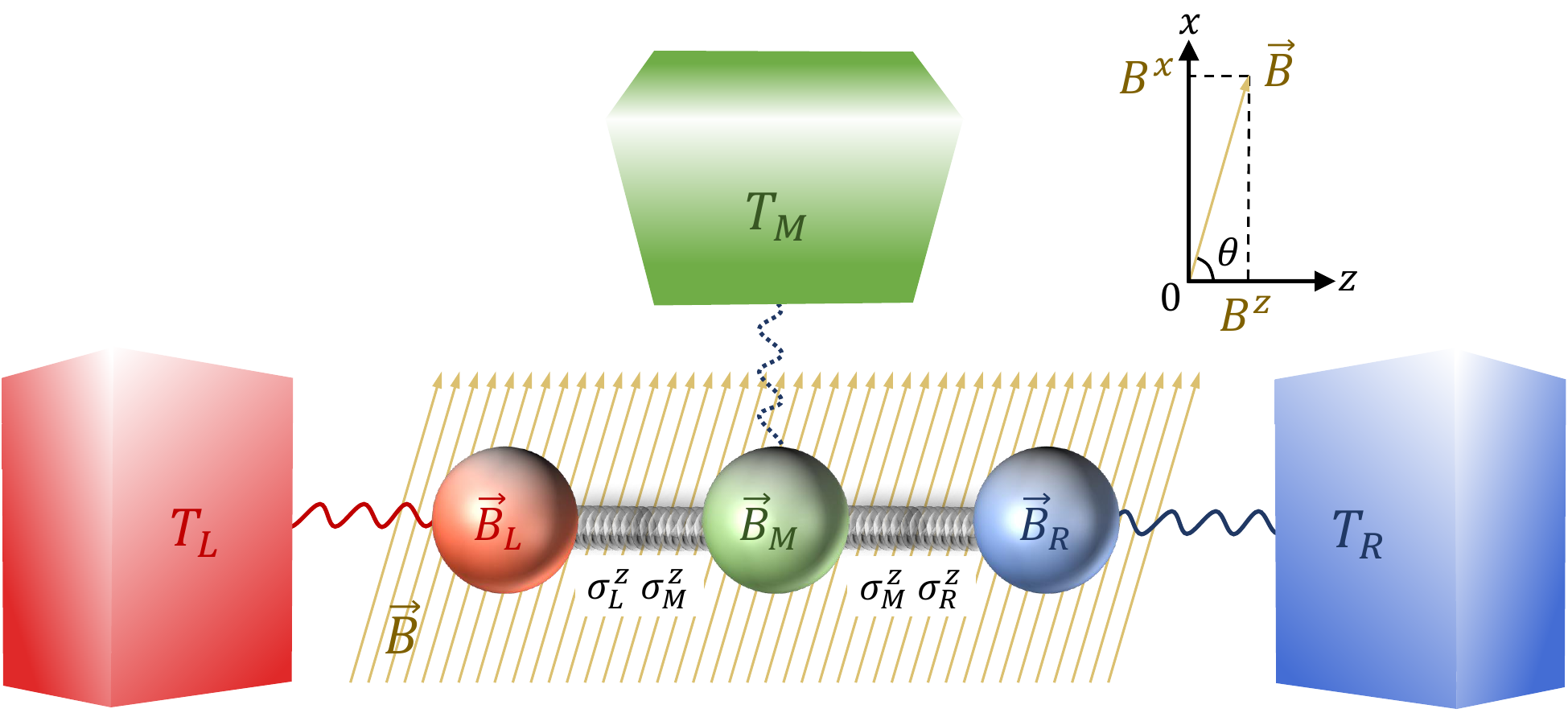}
	\caption{Sketch of the magnetically controlled thermal devices. The system consists of three nearest-neighbor coupled spin-1/2 systems immersed in a direction-adjustable skew magnetic field (SF) $\vec{B}$ with the fixed amplitude $B$, i.e., $\vec{B}=B^x\vec{e}_x+B^z\vec{e}_z$, where $B^x=B\sin\theta$ and $B^z=B\cos\theta$. The coupling type between each two spins is $\sigma_\mu^z\sigma_\nu^z$, $\mu,\nu=L, M, R$. The left, middle, and right atoms are labeled by $L$, $M$, and $R$. The energy of each atom in the magnetic field is $B_\mu$. Each spin is independently connected to a thermal reservoir with the temperature $T_\mu$. The dotted line between the middle spin and the corresponding reservoir indicates that the reservoir $M$ is a perturbing reservoir provided that the system is considered as a two-terminal device.}
\label{model}
\end{figure}
The system, consisting of three nearest-neighbor spin-1/2 systems which are connected to an independent local heat reservoir, immersed in a directional adjustable magnetic field $\vec{B}$ is the model mainly considered in this work, as shown in Fig. \ref{model}.
To simplify, we assume that the magnetic field is restricted in the $x$-$0$-$z$ plane, i.e., $\vec{B}=B^x\vec{e}_x+B^z\vec{e}_z$, where $B^x=B\sin\theta$ (or $B^z=B\cos\theta$) is the component of the magnetic field on the $x$- (or $z$-) axis, $\vec{e}_x$ and $\vec{e}_z$ are unit vectors, $B\equiv\vert\vec{B}\vert=\sqrt{\vert B^z\vert^2+\vert B^x\vert^2}$ represents the constant amplitude of the magnetic field, and $\theta$ represents the angle between the magnetic field and the horizontal direction.
$\theta=s\pi$ or $\theta=\frac{2s+1}{2}\pi$, $s=0,1,2,\cdot\cdot\cdot,$ corresponds to the longitudinal field (LF) or the transverse field (TF), which means that the direction of the magnetic field parallel or perpendicular to the coupled spins.

When the direction of the magnetic field $\theta$ is fixed, the Hamiltonian of the coupled spin-1/2 systems reads
\begin{equation}
H_S=H_{S0}+H_{SI}.
\end{equation} 
The Zeeman energy generated by the coupling of spin and magnetic field $H_{S0}$ is (we take $\hbar=k_B=1$)
\begin{align}
H_{S0}=\frac{1}{2}\sum_\mu(B_\mu^x\sigma_\mu^x+B_\mu^z\sigma_\mu^z),\quad \mu=L,M,R,
\end{align}
where $B_\mu^x=B_\mu\cos\theta$ (and $B_\mu^z=B_\mu\sin\theta$) denotes the TF (and LF) of the $\mu$th spin with $B_\mu$ denoting the energy gap of $\mu$th spin induced by the magnetic field. 
The interaction Hamiltonian $H_{SI}$ between the two nearest-neighbor coupled spins is 
\begin{align}
H_{SI}=\frac{1}{2}(J_{LM}\sigma_L^z\sigma_M^z+J_{MR}\sigma_M^z\sigma_R^z),
\end{align}
where $J_{\mu\nu}$ denotes the interaction strength between the $\mu$th and the $\nu$th spins. In this system, the coupling between the left and right spins is disregarded, i.e., $J_{LR}=0$. $\sigma^x_\mu$ and $\sigma^z_\mu$ are the $x$-axis and $z$-axis component of the spin operator, expressed as
\begin{align}
\sigma_\mu^x=
\begin{pmatrix}
0&1\\1&0
\end{pmatrix},\quad
%\sigma_\mu^y=
%\begin{pmatrix}
%0&-i\\i&0
%\end{pmatrix},\quad
\sigma_\mu^z=
\begin{pmatrix}
1&0\\0&-1
\end{pmatrix}.
\end{align}
This Ising-like spin chain in the skew field has been applied in many scenarios  \cite{NETO20131,nature09994,YUAN2021126279,PhysRevB.68.214406,PhysRevB.85.134412,PhysRevA.103.043312,PhysRevA.106.053308}. For example, this model was simulated by confining the rubidium atoms in the optical lattice and the ferromagnetic-antiferromagnetic quantum phase transition has been observed \cite{nature09994}.

Each spin is in contact with an independent heat reservoir with the temperature $T_\mu$. The Hamiltonian of the three baths read
\begin{align}
H_E=\sum_\mu H_{E}^\mu=\sum_\mu\sum_k\omega_{\mu k}a_{\mu k}^\dagger a_{\mu k},
\end{align}
where $\omega_{\mu k}$ is the frequency of the $k$th Bosonic mode, and $a^\dagger_{\mu k}$ $(a_{\mu k})$ is the creation (annihilation) operator.
The dipole interaction between the spin and the corresponding reservoir is given by
\begin{align}
H_I=\sum_\mu\sum_kf_{\mu k}\sigma_\mu^x(a^\dagger_{\mu k}+a_{\mu k}),
\end{align}
where $f_{\mu k}$ represents the coupling strength between the $\mu$th spin and the $k$th mode.

The evolution of the system is governed by the Born-Markov-secular (BMS) master equation \cite{10.1093/acprof:oso/9780199213900.001.0001}, which can be written in the Schr$\ddot{\mathrm{o}}$dinger picture as
\begin{align}
\dot{\rho}(t)=-i[H_S,\rho(t)]+\sum_\mu\mathcal{L}_\mu[\rho(t)],\label{dissipator}
\end{align}
where the Lindblad dissipator $\mathcal{L}_\mu[\rho(t)]$ for the $\mu$th spin is
\begin{align}
\nonumber
\mathcal{L}_\mu[\rho(t)]&=J_\mu(-\omega^\mu_{ij})[2V^\mu_{ij}\rho (t){V^\mu_{ij}}^\dagger-\{{V^\mu_{ij}}^\dagger V^\mu_{ij},\rho(t)\}]\\
&+J_\mu(+\omega^\mu_{ij})[2{V^\mu_{ij}}^\dagger\rho (t){V^\mu_{ij}}-\{{V^\mu_{ij}V^\mu_{ij}}^\dagger,\rho(t)\}]\label{dissipatormu}
\end{align}
with $J_\mu(\pm\omega_{ij}^\mu)=\pm\kappa_\mu(\omega_{ij}^\mu)n_\mu(\pm\omega_{ij}^\mu)$ denoting the spectrum density. It is worth mentioning that the dissipation rate $\kappa_\mu(\omega_{ij}^\mu)$ between the $\mu$th spin and the mode of frequency $\omega_{ij}^\mu$ in the corresponding reservoir is a quadratic function of the spin-Boson coupling strength $f_\mu(\omega_{ij}^\mu)$, i.e., $\kappa_\mu(\omega_{ij}^\mu)=\pi f_\mu^2(\omega_{ij}^\mu)$ ($\pi$ stems from the integral formula $\int^{+\infty}_0d\tau e^{-i\omega\tau}=\pi\delta(\omega)-i\mathrm{P}\frac{1}{\omega}$, the Cauchy principal value $\mathrm{P}$ in the imaginary part can be ignored as it can be eventually collected to the Lamb shift). For simplicity, we only consider the flat spectrum, i.e., $\kappa_\mu(\omega_{ij}^\mu)=\kappa_\mu$. $n_\mu(\omega_{ij}^\mu)=[\mathrm{exp}(\omega_{ij}^\mu/T_\mu)-1]^{-1}$ is the average photon number. In Eq. (\ref{dissipatormu}), the eigen-operator $V_{ij}^\mu$ and the corresponding eigenfrequency $\omega_{ij}^\mu$ are defined as
\begin{align}
\sum_{\omega_{ij}^\mu}V^\mu_{ij}=\sum_{\omega_{ij}^\mu=\omega_j-\omega_i}\vert i\rangle\langle i \vert\sigma_\mu^x\vert j\rangle\langle j \vert,\label{eigen}
\end{align}
where the eigen-frequency $\omega^\mu_{ij}$ is the transition energy of the transition $\vert i\rangle\leftrightarrow\vert j\rangle$ induced by the reservoir $\mu$, $\omega_i$ and $\vert i\rangle$ are the eigenvalue and eigenstate of the system Hamiltonian, i.e., $H_S=\sum_i\omega_i\vert i\rangle\langle i \vert$. In particular, the eigenvector can be expanded by the bare basis as $\vert i\rangle=\sum_{j=1}^8\Lambda(i,j)\vert \tilde{j}\rangle$, where $\Lambda(i,j)$ is the matrix element in row $i$ and column $j$, and the explicit expression of $ \Lambda(i,j)$ is omitted here. The bare basis is written as
\begin{align}
\nonumber
\vert\tilde{1}\rangle&=\vert\uparrow\uparrow\uparrow\rangle,\quad\vert\tilde{2}\rangle=\vert\uparrow\uparrow\downarrow\rangle,\quad\vert\tilde{3}\rangle&=\vert\uparrow\downarrow\uparrow\rangle,\quad\vert\tilde{4}\rangle=\vert\uparrow\downarrow\downarrow\rangle,\\
\vert\tilde{5}\rangle&=\vert\downarrow\uparrow\uparrow\rangle,\quad\vert\tilde{6}\rangle=\vert\downarrow\uparrow\downarrow\rangle,\quad\vert\tilde{7}\rangle&=\vert\downarrow\downarrow\uparrow\rangle,\quad\vert\tilde{8}\rangle=\vert\downarrow\downarrow\downarrow\rangle,\label{bare_basis_3}
\end{align} 
where $\vert\uparrow\uparrow\uparrow\rangle=\vert\uparrow\rangle_L\otimes\vert\uparrow\rangle_M\otimes\vert\uparrow\rangle_R$, $\sigma^z_\mu\vert\uparrow\rangle_\mu=\vert\uparrow\rangle_\mu$, and $\sigma^z_\mu\vert\downarrow\rangle_\mu=-\vert\downarrow\rangle_\mu$.
Therefore, the eigen-operators can be explicitly given as
\begin{align}
\nonumber
V^L_{ij}&=\sum_{l=1,2,3,4}[\Lambda(i,l)\Lambda(j,l+4)+\Lambda(i,l+4)\Lambda(j,l)]\vert i\rangle\langle j\vert,\\
\nonumber
V^M_{ij}&=\sum_{l=1,2,5,6}[\Lambda(i,l)\Lambda(j,l+2)+\Lambda(i,l+2)\Lambda(j,l)]\vert i\rangle\langle j\vert,\\
V^R_{ij}&=\sum_{l=1,3,5,7}[\Lambda(i,l)\Lambda(j,l+1)+\Lambda(i,l+1)\Lambda(j,l)]\vert i\rangle\langle j\vert.\label{bare-basis}
\end{align}
with $i\in[1,7]$ and $j\in[i+1,8]$. One can find that every heat reservoir can induce the transition between any two eigenvectors $\vert i\rangle$ and $\vert j\rangle$ of the system, but the probability of this transition is different.

Eq. (\ref{dissipator}) can be divided into two completely unrelated groups \cite{PhysRevE.76.031115} corresponds to the evolution of the populations and coherences, i.e., the diagonal and non-diagonal elements of the density matrix. The coherence of the system decays to zero at steady state, so the steady-state density matrix must be diagonal in the $H_S$ representation.
As a consequence, the dynamics of the diagonal entries can be given as 
\begin{align}
\dot{\rho}_{ii}=-\sum_\mu\sum_{j=1,j\neq i}\Gamma_{ij}^\mu,\quad i=1,\cdot\cdot\cdot,8,\label{dynamic_total}
\end{align}
where $\Gamma^\mu_{ij}=2[J_\mu(+\omega_{ij}^\mu){\rho}_{ii}-J_\mu(-\omega_{ij}^\mu)\rho_{jj}]=-\Gamma^\mu_{ji}$ is the net transition rate, and $\rho_{ii}$ denotes the population of the energy level $\vert i\rangle$ in the $H_S$ representation.
The steady state should satisfy 
\begin{align}
\vert\dot{\rho}^S\rangle=\mathcal{M}\vert\rho^S\rangle=0\label{Condsteady}
\end{align}
and the normalization condition $\mathrm{Tr}\rho^S=1$, where $\vert\rho^S\rangle=[\rho_{11}^S,\rho_{22}^S,\rho_{33}^S,\rho_{44}^S,\rho_{55}^S,\rho_{66}^S,\rho_{77}^S,\rho_{88}^S]^T$ and
$\mathcal{M}=\sum_\mu\mathcal{M}_\mu$ is coefficient matrix corresponding to Eq. (\ref{dynamic_total}).
Generally, the rank of the coefficient matrix $\mathcal{M}$ is 7, so the steady state $\vert\rho^S\rangle$ should be unique.
Eq. (\ref{Condsteady}) can be analytically solved, but we don't present it here because the explicit expression is too complicated.
However, if the magnetic field is along some particular direction, the dynamics could become simple, thus the steady state could be explicitly provided.

\textit{Longitudinal field case.-}
The LF corresponds to $B_\mu^z=B_\mu$ and $B_\mu^x=0$. We use $\parallel$ to denote the LF case. For example, the eigenvector of the Hamiltonian $H_S^\parallel$ is $\vert i^\parallel\rangle$ and others are similar. In the LF case, one can find that the corresponding eigenvalues of the Hamiltonian $H_S^\parallel$ are given in Appendix \ref{AppendixA}, and the transformation matrix $\Lambda^\parallel$ is an identity. Therefore, the eigenvectors in the LF case are exactly the bare basis. Meanwhile, the dynamics Eq. (\ref{dynamic_total}) can be simplified as Eq. (\ref{dynamics}). The transitions induced by every reservoir given in Eq. (\ref{bare-basis}) can be reduced to 4, namely, $\vert 1^\parallel\rangle\leftrightarrow\vert 5^\parallel\rangle$, $\vert 2^\parallel\rangle\leftrightarrow\vert 6^\parallel\rangle$, $\vert 3^\parallel\rangle\leftrightarrow\vert 7^\parallel\rangle$, and $\vert 4^\parallel\rangle\leftrightarrow\vert 8^\parallel\rangle$ induced by reservoir $L$, $\vert 1^\parallel\rangle\leftrightarrow\vert 3^\parallel\rangle$, $\vert 2^\parallel\rangle\leftrightarrow\vert 4^\parallel\rangle$, $\vert 5^\parallel\rangle\leftrightarrow\vert 7^\parallel\rangle$, and $\vert 6^\parallel\rangle\leftrightarrow\vert 8^\parallel\rangle$ induced by reservoir $M$, and $\vert 1^\parallel\rangle\leftrightarrow\vert 2^\parallel\rangle$, $\vert 3^\parallel\rangle\leftrightarrow\vert 4^\parallel\rangle$, $\vert 5^\parallel\rangle\leftrightarrow\vert 6^\parallel\rangle$, and $\vert 7^\parallel\rangle\leftrightarrow\vert 8^\parallel\rangle$ induced by reservoir $R$.
The matrix $\mathcal{M}^\parallel$ in the evolution equation $\vert\dot{\rho}^{\parallel}(t)\rangle=\mathcal{M}^\parallel\vert\rho^\parallel(t)\rangle$ is 7, so the steady state $\rho^{\parallel,S}$ is uniquely determined.
Because of its complex expression, the concrete expression of the steady state is not given here.

\textit{LF without middle reservoir.-} If the middle spin is not in contact with an environment, i.e., $\kappa_M=0$,  the corresponding dissipator $\mathcal{L}_M[\tilde{\rho}^{\parallel}(t)]=0$ in Eq. (\ref{dissipator}). Throughout the paper, we use $\tilde{}$ to denote the case $\kappa_M=0$. In this case, the transitions induced by the reservoir $M$ are forbidden, namely, $\tilde{\Gamma}^{\parallel,M}_{ij}=0$ in Eq. (\ref{dynamics}). Based on the allowed transitions, the dynamics of the populations can be divided into two groups, as shown in Eqs. (\ref{rho1256},\ref{rho3478}), the total Hilbert space can be divided into two independent subspaces. The subspace consisting of energy levels $\vert 1^\parallel\rangle$, $\vert 2^\parallel\rangle$, $\vert 5^\parallel\rangle$, and $\vert 6^\parallel\rangle$ is denoted by $S_1^\parallel$ and the subspace supported by $\vert 3^\parallel\rangle$, $\vert 4^\parallel\rangle$, $\vert 7^\parallel\rangle$, and $\vert 8^\parallel\rangle$ is denoted by $S_2^\parallel$.
Since subspaces $S_1^\parallel$ and $S_2^\parallel$ are both subsets of the total space and have no intersection, the Hilbert space of the stable system can be represented as the direct sum of two subspaces, i.e., $S^\parallel=S_1^\parallel\oplus S_2^\parallel$ in the steady state.
In $S_l^\parallel$, $l=1,2$, the evolution equation becomes $\vert\dot{\tilde{\rho}}^\parallel_l(t)\rangle=\tilde{\mathcal{M}}^\parallel_l\vert\tilde{\rho}^\parallel_l(t)\rangle$, where the dissipation coefficient matrix $\tilde{\mathcal{M}}^\parallel_l$ is shown in Appendix \ref{AppendixA}, 
$\vert\tilde{\rho}^\parallel_1(t)\rangle=[\tilde{\rho}^\parallel_{11}(t),\tilde{\rho}^\parallel_{22}(t),\tilde{\rho}^\parallel_{55}(t),\tilde{\rho}^\parallel_{66}(t)]^T$ and $\vert\tilde{\rho}^\parallel_2(t)\rangle=[\tilde{\rho}^\parallel_{33}(t),\tilde{\rho}^\parallel_{44}(t),\tilde{\rho}^\parallel_{77}(t),\tilde{\rho}^\parallel_{88}(t)]^T$. 
%The steady state expressions in two subspaces are also shown in Appendix \ref{AppendixA}. 
Since the evolution of the two subspaces is independent of each other, for a random initial state 
\begin{align}
\nonumber
\tilde{\rho}^{\parallel,I}=&\tilde{p}\sum_{i,j\in S_1^\parallel}\frac{\tilde{\rho}_{ij}^{\parallel,I}}{\tilde{p}}\vert i^\parallel\rangle\langle j^\parallel\vert +(1-\tilde{p})\sum_{i,j \in S_2^\parallel}\frac{\tilde{\rho}_{ij}^{\parallel,I}}{1-\tilde{p}}\vert i^\parallel\rangle\langle j^\parallel\vert \\
+&\sum_{i\in S_1^\parallel,j\in S_2^\parallel}\tilde{\rho}_{ij}^{\parallel,I}\vert i^\parallel\rangle\langle j^\parallel\vert+\sum_{i\in S_2^\parallel,j\in S_1^\parallel}\tilde{\rho}_{ij}^{\parallel,I}\vert i^\parallel\rangle\langle j^\parallel\vert\label{initialstate}
\end{align}
with $\tilde{p}=\sum_{i\in S_1^\parallel}\langle i^\parallel\vert\tilde{\rho}^{\parallel,I}\vert i^\parallel\rangle$ denoting the probability that the system is in the state consisting of eigenvectors of the subspace $S^\parallel_1$ at the initial state, and $\tilde{\rho}_{ij}^{\parallel,I}=\langle i^\parallel\vert\tilde{\rho}^{\parallel,I}\vert j^\parallel\rangle$ representing the matrix element in the $H_S$ representation. In Eq. (\ref{initialstate}), the first (or second) line indicates that the initial state of the system is directly (or interactively) formed by the eigenvectors of the two subspaces. Due to the decoupling of these two subspaces in the steady state, the elements in interactive spaces decay to zero. The steady state can be given as
\begin{align}
\nonumber
\vert\tilde{\rho}^{\parallel,S}\rangle&=\tilde{p}[\tilde{\rho}_{11}^{\parallel,S},\tilde{\rho}_{22}^{\parallel,S},0,0,\tilde{\rho}_{55}^{\parallel,S},\tilde{\rho}_{66}^{\parallel,S},0,0]^T\\
&+(1-\tilde{p})[0,0,\tilde{\rho}_{33}^{\parallel,S},\tilde{\rho}_{44}^{\parallel,S},0,0,\tilde{\rho}_{77}^{\parallel,S},\tilde{\rho}_{88}^{\parallel,S}]^T,
\end{align}
where the explicit expression of $\tilde{\rho}_{ii}^{\parallel,S}$ can be found in Eq. (\ref{rhoSLF}), and the detailed derivation is given in Appendix \ref{AppendixA}.

\begin{figure}
	\centering
		\includegraphics[width=8.3cm]{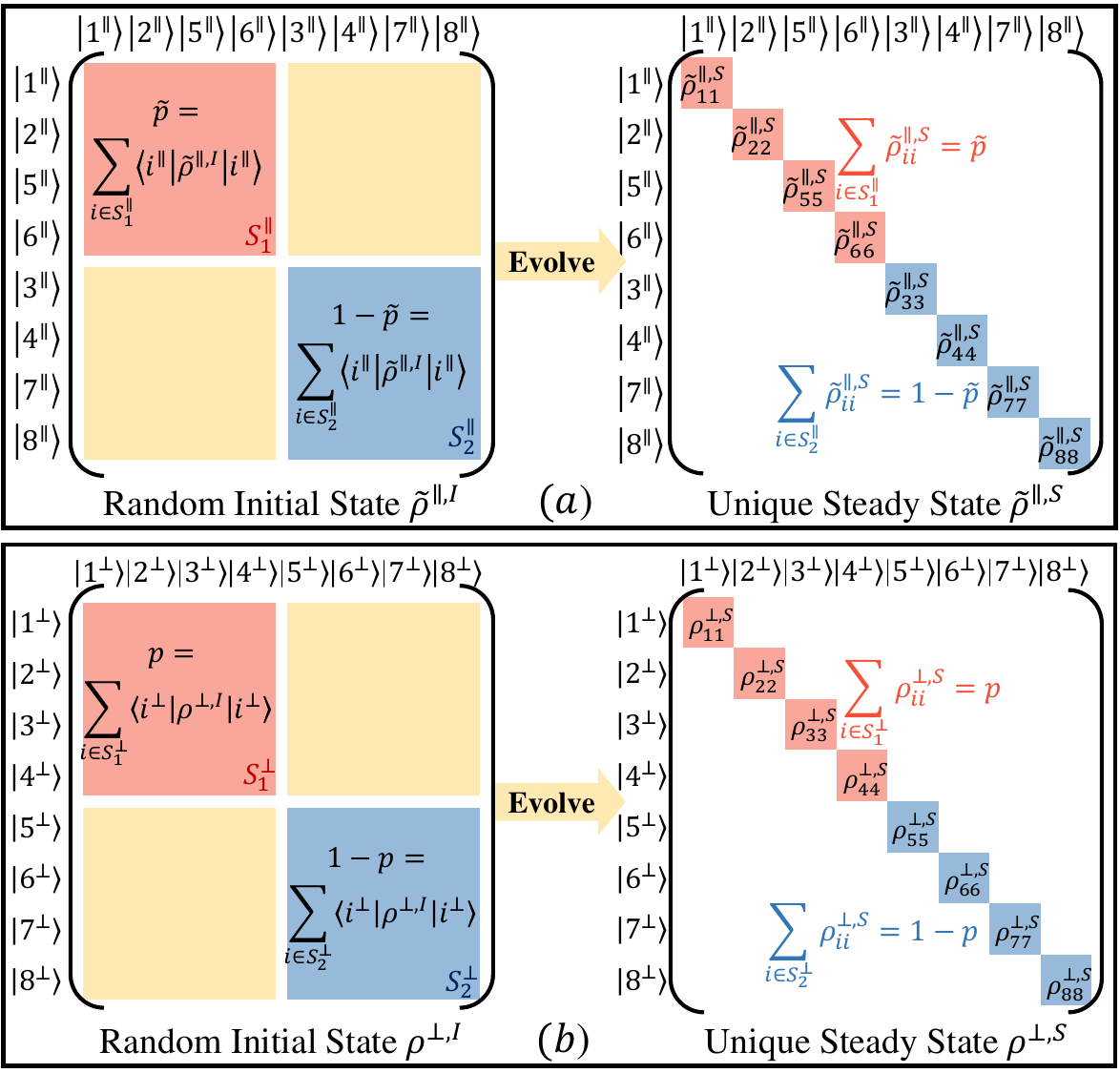}
	\caption{Evolution sketch of density matrix. (a) and (b) correspond to the dynamic processes of the longitudinal field model with $\kappa_M=0$ and the transverse field model with $\kappa_M\neq 0$, respectively. The left or right figure corresponds to the random initial state and unique steady state. The colorful and blank elements in the figure correspond to the appearing and disappearing matrix elements. In the left figure, the red and blue regions correspond to the first and second terms, and the yellow regions correspond to the last two terms in Eq. (\ref{initialstate}). Each element can be represented as $\rho_{ij}=\langle i\vert\rho\vert j\rangle$ for the corresponding density matrices and the eigenstates.}
\label{rho_initial}
\end{figure}

The evolution process of the density matrix is shown in Fig. \ref{rho_initial}(a) in order to describe the dynamics of independent subspaces more clearly. The 8*8 matrix in each parenthesis represents the density matrix, and the colorful and blank elements correspond to the non-zero and zero density matrix elements, respectively. For any given initial state $\tilde{\rho}^{\parallel,I}$, the density matrix elements in the red, blue, and yellow regions correspond to the three terms in Eq. (\ref{initialstate}). When the system is in the steady state, all non-diagonal elements decay to 0, and the component $\tilde{p}$ of the subspace $S_1^\parallel$ is constant, meaning that the subspaces evolve independently of each other.

\textit{Transverse field case.-}
The TF corresponds to $B^x_\mu=B_\mu$ and $B^z_\mu=0$, where we use $\perp$ to specify this case. For example, the eigenvector of $H_S^\perp$ in this case is expressed as $\vert i^\perp\rangle=\sum_{i=1}^8\Lambda^\perp(i,j)\vert\tilde{j}\rangle$ with the transformation matrix correspondingly denoted by $\Lambda^\perp$. 
%Specific derivations of the steady state are given in Appendix \ref{AppendixB}. 
The detailed contents about $\Lambda^\perp$ and the eigenoperators in this case are given in Eqs. (\ref{LambdaHS3},\ref{1234},\ref{5678}).
What we emphasize is that the system can also be divided into two independent subspaces, one consists of the four energy levels $\vert 1^\perp\rangle$, $\vert 2^\perp\rangle$, $\vert 3^\perp\rangle$ and $\vert 4^\perp\rangle$, denoted by $S_1^\perp$, and the other four levels forming the other subspace denoted by $S_2^\perp$. There are six possible transitions in each subspace, the dynamics of the system can be expressed as Eqs. (\ref{rho1234},\ref{rho5678}).
Analogous to Eq. (\ref{initialstate}), if we use $p$ to denote the fraction of the subspaces $S_1^\perp$ in the initial state, the steady state can be expressed as
\begin{align}
\nonumber
\vert\rho^{\perp,S}\rangle&=p[\rho_{11}^{\perp,S},\rho_{22}^{\perp,S},\rho_{33}^{\perp,S},\rho_{44}^{\perp,S},0,0,0,0]^T\\
&+(1-p)[0,0,0,0,\rho_{55}^{\perp,S},\rho_{66}^{\perp,S},\rho_{77}^{\perp,S},\rho_{88}^{\perp,S}]^T,\label{steat}
\end{align}
where $\rho_{ii}^{\perp,S}$ are shown in Eq. (\ref{steady_TF}) and the detailed derivation is given in Appendix \ref{AppendixB}. The corresponding evolution sketch is shown in Fig. \ref{rho_initial}(b).

\section{Steady-state heat current}
\label{section3}
Through the above section, one can obtain the steady state of the system in various cases, allowing us to study the thermodynamic behaviors. The heat current is defined as the first-order derivative of the mean energy of the system with respect to time, i.e.,
\begin{align}
\sum_\mu\dot{Q}_\mu=\frac{d}{dt}\langle H_S\rangle=\sum_\mu\mathrm{Tr}\{H_S\mathcal{L}_\mu[\rho(t)]\},\label{Qdefination}
\end{align}
with $\langle H_S\rangle=\mathrm{Tr}[H_S\rho(t)]$. $\dot{Q}_\mu$ denotes the heat current from the $\mu$th reservoir to the system.  
With the vectorized populations $\vert\rho(t)\rangle$, the steady-state heat current can be expressed as                     
\begin{align}
\dot{Q}_\mu=\langle\omega\vert \mathcal{M}_\mu\vert \rho^S\rangle,\label{Qdefinationde}
\end{align}
where $\vert\omega\rangle$ is the vector consisting of the energy eigenvalues of the system, and $\mathcal{M}_\mu$ is the dissipation coefficient matrix associated with the reservoir $\mu$.
For the steady-state heat currents, $\sum_\mu\dot{Q}_\mu=0$, which is consistent with the first law of thermodynamics. 

According to Eqs. (\ref{Qdefination},\ref{Qdefinationde}), the steady-state heat current can be obtained for the SF, the LF, and the TF, respectively. 
In the SF case, the heat current can be given as
\begin{align}
\dot{Q}_\mu=\sum_{i=1}^7\sum_{j=i+1}^8\omega_{ij}^\mu\Gamma_{ij}^\mu.
\end{align}

In the LF case, the steady-state heat currents for $\kappa_M\neq 0$ read 
\begin{align}
\nonumber
\dot{Q}_L^\parallel&=\omega_{15}^{L,\parallel}\Gamma^{L,\parallel}_{15}+\omega_{26}^{L,\parallel}\Gamma_{26}^{L,\parallel}+\omega_{37}^{L,\parallel}\Gamma_{37}^{L,\parallel}+\omega_{48}^{L,\parallel}\Gamma_{48}^{L,\parallel},\\
\nonumber
\dot{Q}_M^\parallel&=\omega_{13}^{M,\parallel}\Gamma^{M,\parallel}_{13}+\omega_{24}^{M,\parallel}\Gamma_{24}^{M,\parallel}+\omega_{57}^{M,\parallel}\Gamma_{57}^{M,\parallel}+\omega_{68}^{M,\parallel}\Gamma_{68}^{M,\parallel},\\
\dot{Q}_R^\parallel&=\omega_{12}^{R,\parallel}\Gamma^{R,\parallel}_{12}+\omega_{56}^{R,\parallel}\Gamma_{56}^{R,\parallel}+\omega_{34}^{R,\parallel}\Gamma_{34}^{R,\parallel}+\omega_{78}^{R,\parallel}\Gamma_{78}^{R,\parallel}.
\end{align}
However, if $\kappa_M=0$, it is easy to obtain that any net transition rate $\tilde{\Gamma}^{\mu,\parallel}_{ij}=0$. Although the steady state consists of two independent subspaces, there is no heat transfer in either subspace between the remaining two reservoirs. Therefore, the total steady-state heat currents of the system vanish, i.e.,
\begin{align}
\dot{\tilde{Q}}^\parallel_L=\dot{\tilde{Q}}^\parallel_M=\dot{\tilde{Q}}^\parallel_R=0.\label{hcc}
\end{align}
The detailed derivation is given in Eq. (\ref{A10}).
\begin{figure}
	\centering
		\includegraphics[width=8.3cm]{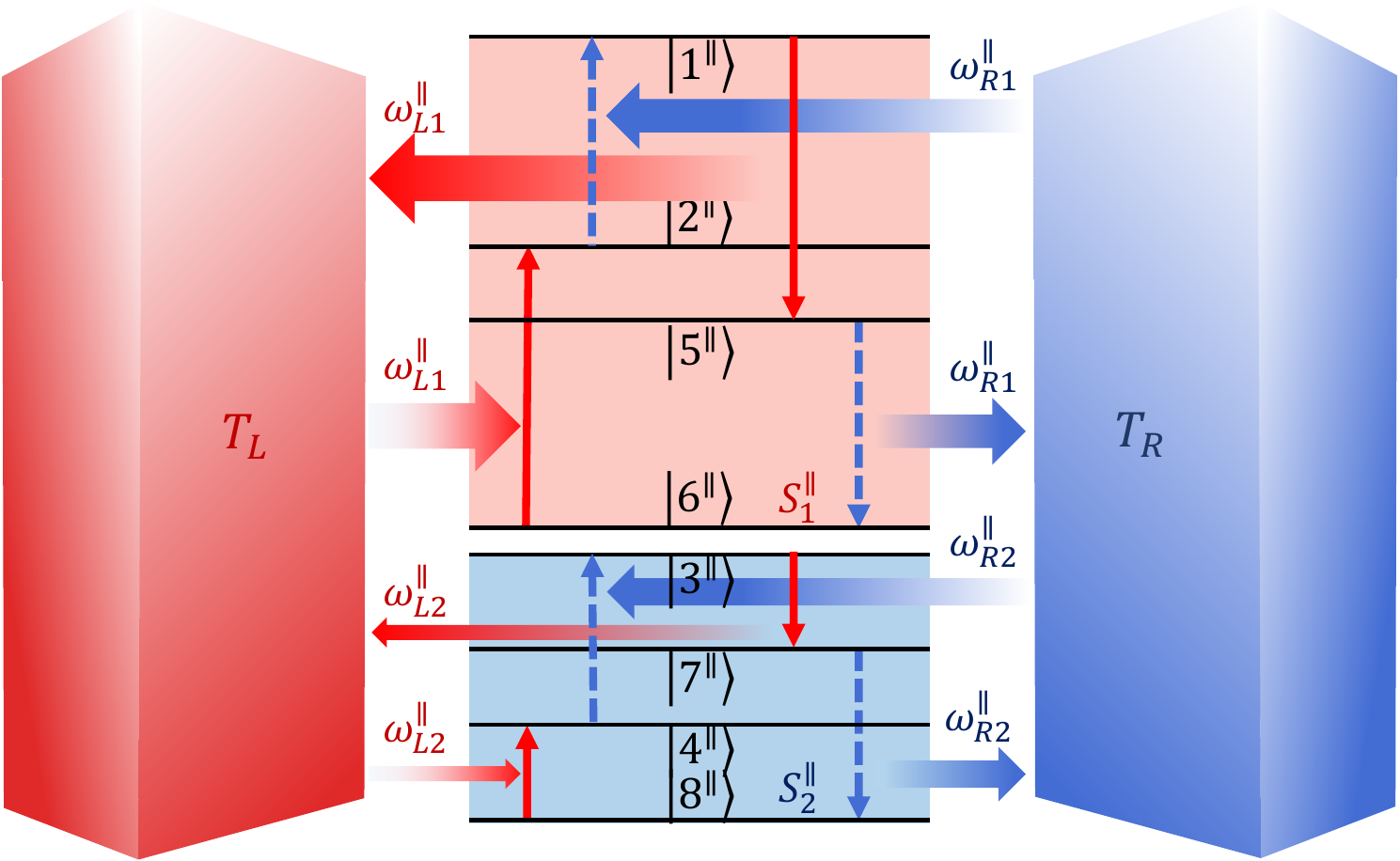}
	\caption{Sketch of the energy level transition of the system when only the longitudinal field is present and $\kappa_M=0$. Since the middle reservoir is not in contact with the system, only two heat reservoirs exist. The red (or blue) shade contains four energy levels that form the subspace $S_1^\parallel$ (or $S_2^\parallel$), respectively. The gradient arrows between the reservoir and the system indicate the energy absorbed or released from the environment, measured by the arrow thickness. The arrows between the energy levels represent stimulated absorption (or spontaneous radiation) that occurs when energy is absorbed from (or released to) the reservoir, and the solid red and dashed blue arrows correspond to transitions induced by the left and right reservoirs. Here, $B_\mu=B_0$, $J_{LM}=0.5B_0$, $J_{MR}=0.1B_0$, $T_L>T_R$ and $\kappa_L=\kappa_R$.}
\label{Qequ0}
\end{figure}

A potential understanding of the blockade of heat current for $\kappa_M=0$ is as follows. In this case, the eigen-operators of the system can be further expressed as Eq. (\ref{V_LR}), and the transitions induced by the middle reservoir are prohibited. The allowable transitions for left (or right) spin can be expressed as the tensor product of the transition of spin itself $\sigma_{L(R)}^-$ with the state of the middle spin $\vert\uparrow\rangle_M\langle\uparrow\vert$ or $\vert\downarrow\rangle_M\langle\downarrow\vert$. 
Both transitions induced by the left (or right) reservoir correspond to the same frequency $\omega_{Ll}$ (or $\omega_{Rl}$) in subspace $S_l^\parallel$. It is well-known that the heat transfer between the system and the environment is generated by the cyclic transition between the energy levels induced by the heat reservoir. 
Fig. \ref{Qequ0} gives the transitions in subspace $S^\parallel_1$ (or $S^\parallel_2$) with the red (or blue) background. For $S_1^\parallel$, we find that there is only one non-repetitive cyclic process inducing the heat transfer: $\vert 6^\parallel\rangle\rightarrow\vert 2^\parallel\rangle\rightarrow\vert 1^\parallel\rangle\rightarrow\vert 5^\parallel\rangle\rightarrow\vert 6^\parallel\rangle$. In this circle, the energy $\omega_{L1}$ absorbed from the high-temperature thermal reservoir (left reservoir) causes the transition $\vert 6^\parallel\rangle\rightarrow\vert 2^\parallel\rangle$, and the equivalent energy $\omega_{L1}$ released to the environment during the transition $\vert 1^\parallel\rangle\rightarrow\vert 5^\parallel\rangle$. Similarly, the energy absorbed and released for right reservoir induced transitions $\vert 2^\parallel\rangle\rightarrow\vert 1^\parallel\rangle$ and $\vert 5^\parallel\rangle\rightarrow\vert 6^\parallel\rangle$ is both $\omega_{R1}$. Therefore, energy cannot be transferred from the left reservoir to the right one in subspace $S_1^\parallel$. 
In the same way, there is no net heat transfer through the subspace $S_2^\parallel$ between two heat reservoirs with temperature gradients. What's more, we also find that the various coupling types between spins do not block heat current except $\sigma_z^\mu\sigma_z^\nu$ coupling, in the dissipative model, Appendix \ref{AppendixC} complements the heat current modulation process when the Heisenberg interaction is considered between the nearest-neighbor spins. 

In the TF case, the steady state of the system is the combination of those in two independent subspaces as given in Eq. (\ref{steat}). The steady-state heat current is also the combination of the contributions of the two subspaces. Namely, the steady-state heat current is determined by the initial state and can be given as
\begin{align}
\dot{Q}_\mu^\perp&=p\sum_{i=1}^3\sum_{j=1+1}^4\omega_{ij}^{\mu,\perp}\Gamma_{ij}^{\mu,\perp}+(1-p)\sum_{i=5}^7\sum_{j=1+1}^8\omega_{ij}^{\mu,\perp}\Gamma_{ij}^{\mu,\perp}.\label{Q_TF}
\end{align}

\begin{figure*}
	\centering
    \begin{subfigure}
		\centering
		\includegraphics[width=18cm]{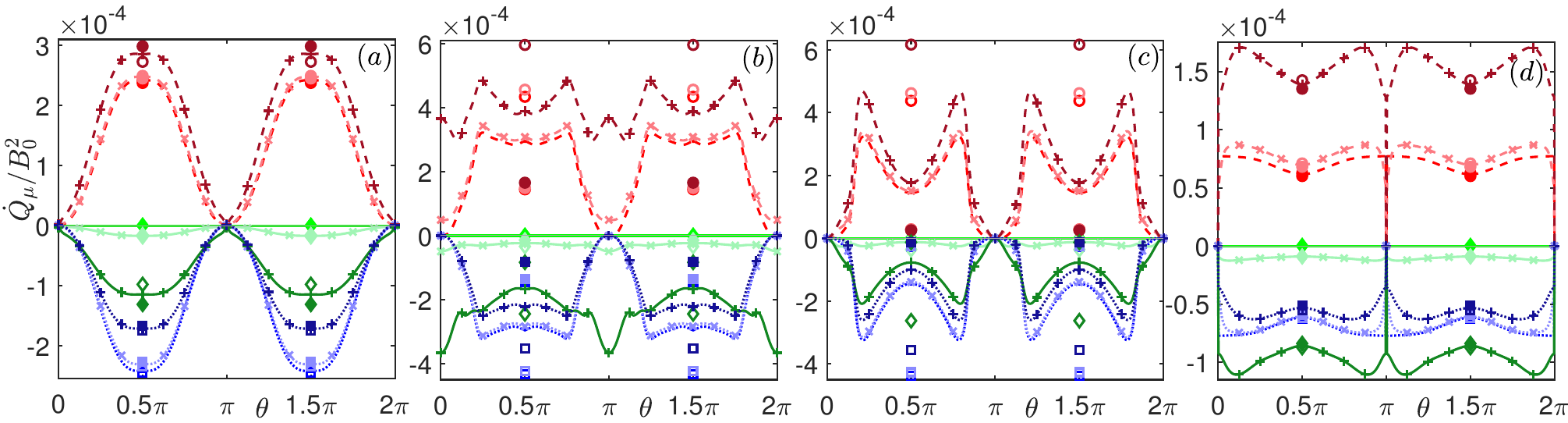}
	\end{subfigure}	
    \begin{subfigure}
		\centering
		\includegraphics[width=18cm]{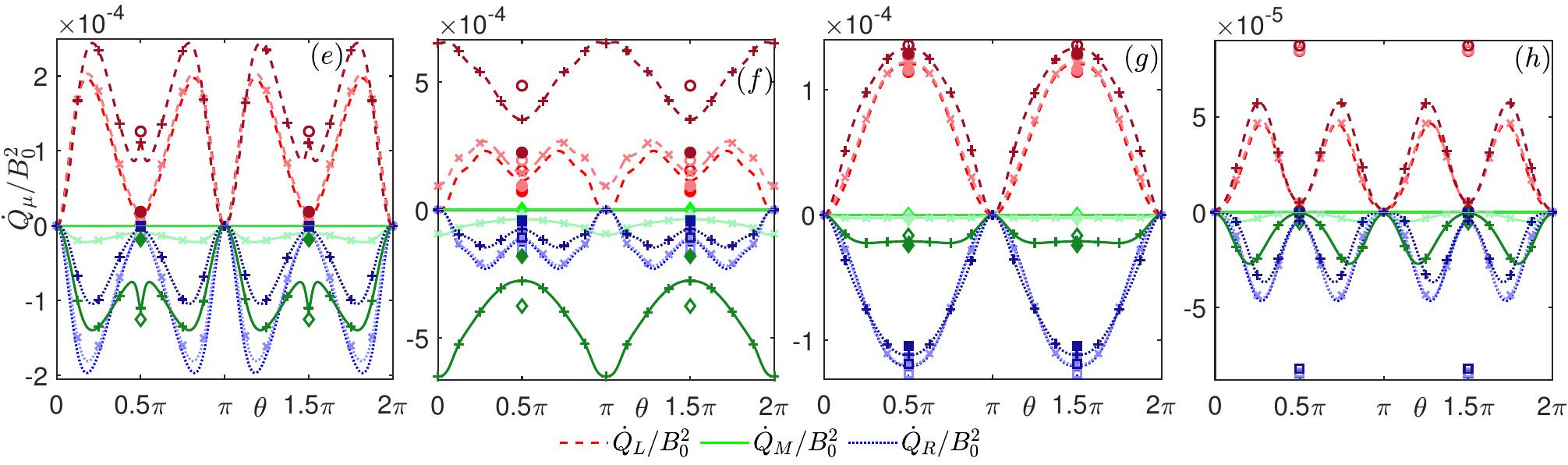}
	\end{subfigure}	
	\caption{Steady-state heat currents with the direction of the magnetic field $\theta$. The red dashed, green solid, and blue dotted lines are used to denote the heat current $\dot{Q}_L$, $\dot{Q}_M$, and $\dot{Q}_R$, the hollow (or solid) circles, diamonds, and squares represent the steady-state heat currents $\dot{Q}_L$, $\dot{Q}_M$, and $\dot{Q}_R$ with $p=0$ (or $p=1$) in the TF case, respectively. Normal-colored unmarked, light-colored with crosses, and dark-colored with plus signs lines describe the heat currents when the dissipation between the middle spin and the environment is $\kappa_M=0$, $\kappa_M=0.1\kappa_L$, and $\kappa_M=\kappa_L$, respectively. (a)-(c) analyze the effect of the amplitude of the magnetic field on the heat current when the coupling strengths between spins are fixed as $J_{LM}=0.8B_0$, $J_{MR}=1.2B_0$. The magnetic fields are considered to be the same, i.e., $B_L=B_M=B_R=B$, $B=0.3B_0$, $B=1.5B_0$ and $B=3B_0$ in (a), (b), and (c). (e)-(g) study the effect of coupling strength between spins on the heat current when the magnetic fields are fixed as $B_L=B_0$, $B_M=2B_0$, and $B_R=3B_0$. The coupling strengths are assumed to be the same, i.e., $J_{LM}=J_{MR}=J$, $J=0.5B_0$, $J=2B_0$, and $J=8B_0$ in (e), (f), and (g). (d) and (h) describe the heat currents when the magnetic field and coupling strength are the same, i.e., $B_L=B_M=B_R=J_{LM}=J_{MR}=B$. $B=0.1B_0$ and $B=10B_0$ in (d) and (h). Other parameters in all the figures are taken as $B_0=1$, $\kappa_L=\kappa_R=0.001B_0$, $T_L=2B_0$, $T_M=0.02B_0$, and $T_R=0.2B_0$.}
\label{Q_theta}
\end{figure*}

\section{Magnetically controlled thermal devices}
\label{section4}
\subsection{Quantum thermal modulator}
\label{section4A}
Let's first consider a two-terminal thermal device, namely, the dissipation of the middle spin does not exist, i.e., $\kappa_M=0$. In this case, the system can be decoupled by the LF ($\theta=s\pi$) into two subspaces where the steady-state heat currents are both zero. 
But the steady-state heat current does not vanish for $\theta\neq s\pi$. This property allows the steady-state heat current to be modulated by controlling the direction of the magnetic field $\theta$. 
We plot the heat current versus the magnetic field direction $\theta$ in Fig. \ref{Q_theta} to demonstrate a whole modulation.

In Fig. \ref{Q_theta}, the dashed red, solid green, and dotted blue lines denote $\dot{Q}_L$, $\dot{Q}_M$, and $\dot{Q}_R$, respectively. Let's focus on the \textit{unmarked lines}. Since we do not consider the middle reservoir ($\kappa_M=0$), one can easily find that the dotted green line shows the zero heat current $\dot{Q}_M=0$. 
From the dashed red and dotted blue lines, one can find that the heat currents $\dot{Q}_L=-\dot{Q}_R$ disappear at $\theta=s\pi$, which is consistent with our above calculation in Eq. (\ref{hcc}), and appear at $\theta\neq s\pi$. This phenomenon illustrates the apparent modulation of the heat current by controlling the magnetic field direction $\theta$. In one modulation period, the magnetic field changes from the LF to the TF and then back to the LF. However, it can be found that the steady-state heat current does not always increase as the magnetic field shifts from the LF to the TF, which closely depends on $B_\mu$ and the coupling $J_{\mu\nu}$. Several typical cases can be observed more clearly by comparing the figures in Figs. \ref{Q_theta}. From Figs. \ref{Q_theta} (a-c), one can find that the maximum heat current occurs far away from the $\theta=(2s+1)\pi/2$ with the magnetic field $B$ increases. And from Figs. \ref{Q_theta} (e-g), one can find that the maximal heat currents occur at $\theta=(2s+1)\pi/2$ for strong coupling $J_{\mu\nu}$ as shown in Fig. \ref{Q_theta}(g) and deviate from $(2s+1)\pi/2$ for the weak and medium coupling as shown in Fig. \ref{Q_theta}(e) and (f). In one word, the modulator can be designed on purpose based on the practical scenario.

Suppose the quantum thermal modulator is perturbed by a perturbing reservoir, i.e., $\kappa_M\neq 0$, the steady-state heat current versus the direction of magnetic field $\theta$ for $\kappa_M=0.1\kappa_L$ and $\kappa_M=\kappa_L$ are depicted in Figs. \ref{Q_theta} by \textit{cross lines} and \textit{plus lines}, respectively. It is clear that when the system is uncoupled ($\kappa_M=0$) or ultra-weakly coupled ($\kappa_M=0.1\kappa_L$) to the middle thermal reservoir, the system can realize the good magnetically controlled modulation of heat current. For $\kappa_M=\kappa_L$, the plus lines in Fig. \ref{Q_theta}(b) show that the heat current changes in a quite limited range with the change of $\theta$, so it is not sufficient to serve as an efficient thermal modulator in the case of the medium magnetic field. In this sense, if the perturbing reservoir leads to strong dissipation (large $\kappa_M$) in the corresponding parameter condition, the magnetically controlled modulator can only be realized at the expense of the weak magnetic field, such as the case in Fig. \ref{Q_theta}(a). On the contrary, if we consider other parameters, as shown in Fig. \ref{Q_theta}(f), one can find that the strong coupling with a proper magnetic field can also serve as a good modulator as shown in Fig. \ref{Q_theta}(g) — the other figures in Fig. \ref{Q_theta} illustrate alternative parameter conditions to achieve the potential modulation effect.

It is exciting that if $\kappa_M=\kappa_L$, one can find that applying the LF can also block the steady-state heat current when $B_\mu=J_{\mu\nu}$. The analytic proof is given in Appendix \ref{AppendixA}. The non-vanishing heat current will appear when the magnetic field direction deviates from the LF. One can also modulate the heat current by controlling the magnetic field. In Figs. \ref{Q_theta}(d) and (h), we show the magnetic modulation process under the condition $B_\mu=J_{\mu\nu}=B$. It indicates that as $B$ increases, the modulation capability of the system improves.

In Fig. \ref{Q_theta}(b), (c), (d), (f), and (h), one can easily observe that there is $\dot{Q}_M >\dot{Q}_R$ near $\theta=s\pi$ when $\kappa_M=\kappa_L$. This is due to the special energy level structure of the system in the condition of $B_\mu\sim J_{\mu\nu}$. With conditions $B_\mu=J_{\mu\nu}$ and $\theta=s\pi$, the levels $\vert 3^\parallel\rangle$, $\vert 4^\parallel\rangle$, $\vert 7^\parallel\rangle$, and $\vert 8^\parallel\rangle$ are  degenerated, so that the transitions $\vert 3^\parallel\rangle\leftrightarrow\vert 7^\parallel\rangle$ and $\vert 4^\parallel\rangle\leftrightarrow\vert 8^\parallel\rangle$ induced by left reservoir (and $\vert 3^\parallel\rangle\leftrightarrow\vert 4^\parallel\rangle$ and $\vert 7^\parallel\rangle\leftrightarrow\vert 8^\parallel\rangle$ induced by the right one) are prohibited, while the transitions induced by the middle reservoir are not affected, so that the heat current from the left reservoir to the middle one is easier than to the right one. 
Of course, the heat transfer between the system and the environment depends not only on the energy level transition but also on the reservoir's temperature. In Appendix \ref{AppendixD}, we give the variations of heat currents with the temperature of the middle heat reservoir when $\theta=0.1\pi$, the parameters are the same as those in Fig. \ref{Q_theta}. There is a clear competition between $\dot{Q}_M$ and $\dot{Q}_R$, and the direction of $\dot{Q}_M$ changes when $T_M$ reaches a critical value.

\begin{figure}
	\centering
		\includegraphics[width=8.3cm]{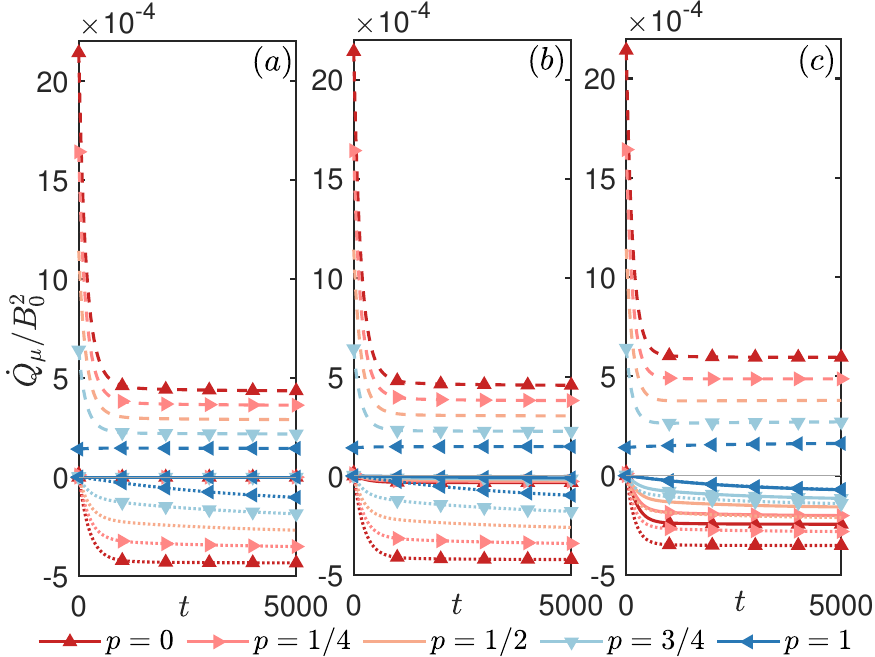}
	\caption{Transient-state heat current with time $t$. The dashed, solid, and dotted lines denote the heat current $\dot{Q}_L$, $\dot{Q}_M$, and $\dot{Q}_R$. The same color corresponds to the same fraction $p$.
Other parameters in all the figures are taken as $B_0=1$, $J_{LM}=0.8B_0$, $J_{MR}=1.2B_0$, $\kappa_L=\kappa_R=0.001B_0$, $T_L=2B_0$, $T_M=0.02B_0$, $T_R=0.2B_0$, $B_L=B_M=B_R=1.5B_0$, $\kappa_M=0$, $\kappa_M=0.1\kappa_L$, and $\kappa_M=\kappa_L$ in (a), (b), and (c).}
\label{Q_time}
\end{figure}

Moreover, as mentioned previously, when $\theta=\frac{2s+1}{2}\pi$ (corresponding to the TF), the steady state of the system depends on the initial state, so do the steady-state heat currents, which should be the joint contribution of the two separate subspaces. In Fig. \ref{Q_theta}, we use circles, diamonds, squares to mark $\dot{Q}_L$, $\dot{Q}_M$, and $\dot{Q}_R$ at $\theta=\frac{2s+1}{2}\pi$ with the 'solid' and 'hollow' signs representing $p=0$ and $p=1$, respectively. Obviously, the difference between the heat currents of the two subspaces decoupled by the TF increases as $B$ increases, which is concretely represented in Figs. \ref{Q_theta}(c), (e), and (h). It must be emphasized that the heat current values at $\theta=\frac{2s+1}{2}\pi$ on all the lines in Fig. \ref{Q_theta} are plotted by choosing the steady state at the almost nearest neighbor point before $\theta=\frac{2s+1}{2}\pi$ as the initial state. This means that the heat current must lie between the solid and hollow signs, regardless of the initial state. However, for a given initial state, the steady-state heat current is uniquely determined. This is quite consistent with the practical scenario where during the adjustment of $\theta$, the initial state of the system for a given $\theta$ is just the steady state of the system corresponding to the $\theta$ before the adjustment.
Of course, one can simultaneously choose some particular initial state by which one could reach different steady states in the TF case. However, due to the non-vanishing heat currents in either subspace for the TF case, the total heat current cannot be zero by changing the fraction $p$, which is explicitly shown in Fig. \ref{Q_time}. The variations of the transient-state heat currents with time $t$ at $\kappa_M=0$, $\kappa_M=0.1\kappa_L$, and $\kappa_M=\kappa_L$ for different $p$ are shown successively in Figs. \ref{Q_time}(a), (b), and (c). The initial state for Fig. \ref{Q_time} is selected as $\vert\rho^{\perp,I}\rangle=p\vert\rho^{\perp,I}_1\rangle+(1-p)\vert\rho^{\perp,I}_2\rangle$ with $\vert\rho^{\perp,I}_1\rangle=[1,0,0,0,0,0,0,0]^T$ and $\vert\rho^{\perp,I}_2\rangle=[0,0,0,0,1,0,0,0]^T$.

In summary, the model can be used as a magnetically controlled thermal modulator, regardless of whether the middle spin is in contact with the environment. When there is no dissipation of the middle spin, the system can perfectly modulate the heat current, i.e., the heat current can be modulated from zero to a finite value. If the modulation of the heat current by the magnetic field is insufficient, it is desirable to control the heat current based on the initial state when the magnetic field is significant.
\begin{figure}
	\centering
		\includegraphics[width=8.3cm]{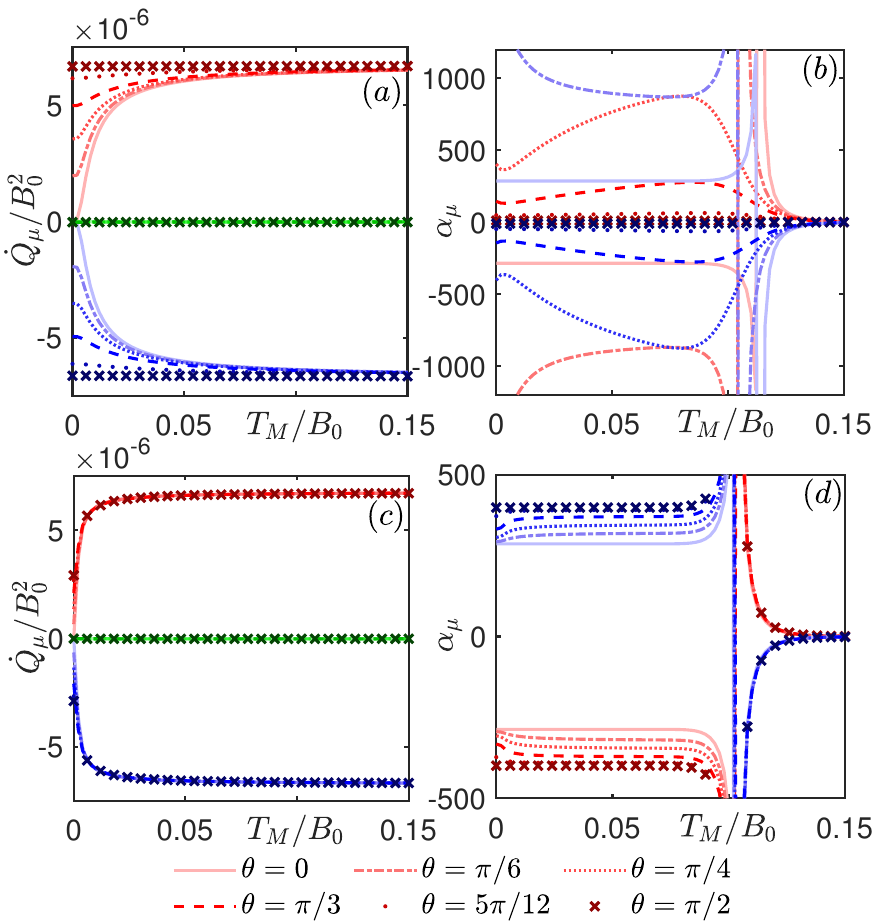}
	\caption{Steady-state heat currents $\dot{Q}_\mu$, $\mu=L,M,R,$ and amplification factor $\alpha_\mu$ versus temperature $T_M$. (a) and (c) describe the steady-state heat currents for different magnetic field directions; (b) and (d) show the amplification factors. In (a) and (c), the red, green, and blue lines denote $\dot{Q}_L$, $\dot{Q}_M$, and $\dot{Q}_R$, respectively. In (b) and (d), the red and blue lines correspond to amplification factors $\alpha_L$ and $\alpha_R$, respectively. Different line styles represent different directions of the magnetic field. Solid lines represent the particular LF case, and cross lines denote the TF case. In (a) and (b), $B_L=B_M=B_R=0.01B_0$, $J_{MR}=1.001B_0$. In (c) and (d), $B_L=B_M=B_R=0.001B_0$, $J_{MR}=1.01B_0$. Other parameters in all the figures are taken as $B_0=1$, $J_{LM}=B_0$, $\kappa_L=\kappa_M=\kappa_R=0.001B_0$, $T_L=0.2B_0$, and $T_R=0.02B_0$.}
\label{alpha}
\end{figure}

\subsection{Quantum thermal transistor}
\label{section4B}
Besides the function of a modulator of heat current, this system can also be used as a well-performing quantum heat transistor or triode, which allows a slight change of the current at one of the terminals (base) to cause a significant change of the currents at the other two terminals (collector and emitter). The amplification characteristics of this system in the LF have been studied extensively \cite{PhysRevLett.116.200601, PhysRevA.97.052112, PhysRevB.101.245402, PhysRevA.103.052613, PhysRevE.106.024110}. 
The performance of a transistor can be described as the amplification factor
\begin{align}
\alpha_\mu=\frac{\partial\dot{Q}_\mu}{\partial\dot{Q}_M},\quad \mu=L,R,
\end{align}       
where the terminal $M$ is regarded as the base. According to the conservation of energy, i.e., $\sum_\mu\dot{Q}_\mu=0$, it is easy to understand that $\alpha_L+\alpha_R=-1$. When $\vert\alpha_\mu\vert>1$, the system can be regarded as a transistor.  The higher the absolute value of $\alpha_\mu$, the better the transistor performance.
Now, we discuss the rectification of the system in the SF and the TF.

Figure \ref{alpha} depicts the variations of the heat currents at three terminals and the amplification factors with the temperature of the base $T_M$ in the different directions of the magnetic field. The heat currents are increasing as $\theta$ increases, as shown in Figs. \ref{alpha}(a), but Fig. \ref{alpha}(b) indicates that the amplification factor is not monotonically changed with $\theta$. In Fig. \ref{alpha}(b), the system achieves good amplification for $\theta=0$ (solid lines), which is consistent with Ref. \cite{PhysRevLett.116.200601}. In addition, Fig. \ref{alpha}(b) also indicates that by adjusting the magnetic field, we can achieve a much larger amplification factor than that in Ref. \cite{PhysRevLett.116.200601}, which can be seen when $\theta=\pi/6$ or $\theta=\pi/4$. Whereas, inappropriate $\theta$ can also lead to the reduction of the amplification factor. For example, the heat currents reach maximum for $\theta=\frac{\pi}{2}$, but the amplification factor is relatively small, $\vert\alpha_\mu\vert\approx 10$. 
However, the amplification factor is not always small in the TF case ($\theta=\pi/2$). When the magnetic field $B_\mu$ is reduced, as the magnetic field gradually changes from $\theta=0$ to $\theta=\frac{\pi}{2}$, both heat currents and amplification factor gradually increase, and the system in the TF can still be treated as a good transistor, which is shown in Figs. \ref{alpha}(c) and (d). 

\section{Discussions and Conclusions}
\label{section5}
Before the end, we would like to discuss the nonreciprocal heat transport through our system to give a comprehensive understanding. We know that the unidirectional thermal transport of the quantum thermal diode usually arises from the asymmetry of the system, such as the different magnetic fields on the spins or the different couplings between the spins in our model. Since there is no heat transport for $\theta=s\pi$, we only address the case $\theta\neq s\pi$. We found that the perfect rectification can be realized for one zero-temperature terminal. At other parameter regimes, our model also demonstrates certain rectification functions. Significantly, the rectification due to different magnetic fields is superior to rectification due to different coupling strengths between spins. Detailed studies are provided in Appendix \ref{AppendixE}.
For the case where the steady state depends on the initial state, although the system is decoupled into two subsystems, the steady-state heat currents in both subsystems are not zero.
Hence, the rectification is closely related to the fraction $p$ of the initial state in the two subspaces, which differs from the case in Ref. \cite{PhysRevE.107.064125} where the rectification is independent of the fraction.

In conclusion, we have designed a quantum thermal device based on three nearest-neighbor coupled spins to change the heat current by changing the direction of the magnetic field. When the middle spin is not connected to the thermal reservoir, we find that the LF can block the heat transfer between the system and the thermal reservoir. With the direction of the magnetic field changed, our system can act as an ideal quantum thermal modulator to manage the heat current by controlling the direction of the magnetic field. A well-performing thermal modulator can also be implemented even if the middle spin is connected to a perturbation reservoir. In particular, the heat current can always be entirely blocked by appropriately adjusting the magnetic field and the coupling strengths between the spins. 
In the TF, we find that the system decouples into two independent subspaces, which provides an alternative modulation method of heat current by controlling the fractions of the initial state in both subspaces. 
When the system serves as a transistor, we show that the controllable magnetic field can improve the performance of the transistor compared to the LF case. This discovery opens up exciting possibilities for improving quantum thermal management in various applications.

\section*{Acknowledgements}
This work was supported by the National Natural Science Foundation of China under Grant No.12175029.

\appendix
\section{Detailed derivation of the steady state in LF}
\label{AppendixA}
Provided that the system is immersed in the LF, the Hamiltonian of the system is
\begin{align}
H_S^\parallel=\frac{1}{2}(\sum_\mu B_\mu\sigma^z_\mu+\sum_{\mu\neq \nu}J_{\mu\nu}\sigma_\mu^z\sigma_\nu^z),\quad\mu,\nu=L,M,R.\label{HS3LF}
\end{align}
In the diagonal representation, i.e., $H_{S}^\parallel=\sum_{i=1}^8\omega^\parallel_i\vert i^\parallel\rangle\langle i^\parallel\vert$, the eigenstates are $\vert i^\parallel\rangle=\vert\tilde{i}\rangle$, which are shown in Eq. (\ref{bare_basis_3}), and the corresponding eigenvalues $\omega^\parallel_i$ are
\begin{align}
\nonumber
\omega^\parallel_1=\frac{1}{2}(+B_L+B_M+B_R+J_{LM}+J_{MR}),\\
\nonumber
\omega^\parallel_2=\frac{1}{2}(+B_L+B_M-B_R+J_{LM}-J_{MR}),\\
\nonumber
\omega^\parallel_3=\frac{1}{2}(+B_L-B_M+B_R-J_{LM}-J_{MR}),\\
\nonumber
\omega^\parallel_4=\frac{1}{2}(+B_L-B_M-B_R-J_{LM}+J_{MR}),\\
\nonumber
\omega^\parallel_5=\frac{1}{2}(-B_L+B_M+B_R-J_{LM}+J_{MR}),\\
\nonumber
\omega^\parallel_6=\frac{1}{2}(-B_L+B_M-B_R-J_{LM}-J_{MR}),\\
\nonumber
\omega^\parallel_7=\frac{1}{2}(-B_L-B_M+B_R+J_{LM}-J_{MR}),\\
\omega^\parallel_8=\frac{1}{2}(-B_L-B_M-B_R+J_{LM}+J_{MR}).
\end{align}
The eigen-operators are 
\begin{alignat}{2}
\nonumber
V^\parallel_{L1}&=V_{51}^{L,\parallel}+V_{62}^{L,\parallel},\quad &\omega^\parallel_{L1}&\equiv\omega^{L,\parallel}_{51}=\omega^{L,\parallel}_{62}=B_L+J_{LM},\\
\nonumber
V^\parallel_{L2}&=V_{73}^{L,\parallel}+V_{84}^{L,\parallel},\quad &\omega^\parallel_{L2}&\equiv\omega^{L,\parallel}_{73}=\omega^{L,\parallel}_{84}=B_L-J_{LM},\\
\nonumber
V^\parallel_{M1}&=V_{31}^{M,\parallel},\quad &\omega^\parallel_{M1}&\equiv\omega^{M,\parallel}_{31}=B_M+J_{LM}+J_{MR},\\
\nonumber
V^\parallel_{M2}&=V_{42}^{M,\parallel},\quad &\omega^\parallel_{M2}&\equiv\omega^{M,\parallel}_{42}=B_M+J_{LM}-J_{MR},\\
\nonumber
V^\parallel_{M3}&=V_{75}^{M,\parallel},\quad &\omega^\parallel_{M3}&\equiv\omega^{M,\parallel}_{75}=B_M-J_{LM}+J_{MR},\\
\nonumber
V^\parallel_{M4}&=V_{86}^{M,\parallel},\quad &\omega^\parallel_{M4}&\equiv\omega^{M,\parallel}_{86}=B_M-J_{LM}-J_{MR},\\
\nonumber
V^\parallel_{R1}&=V_{21}^{R,\parallel}+V_{65}^{R,\parallel},\quad &\omega^\parallel_{R1}&\equiv\omega^{R,\parallel}_{21}=\omega^{R,\parallel}_{65}=B_R+J_{MR},\\
V^\parallel_{R2}&=V_{43}^{R,\parallel}+V_{87}^{R,\parallel},\quad &\omega^\parallel_{R2}&\equiv\omega^{R,\parallel}_{43}=\omega^{R,\parallel}_{87}=B_R-J_{MR},\label{V^parallel}
\end{alignat}
where $V^{\mu,\parallel}_{ij}=\vert i^\parallel\rangle\langle j^\parallel\vert$, and $\omega_{ij}^{\mu,\parallel}$ is defined as Eq. (\ref{eigen}). $V^\parallel_{\mu i}$ denotes the $i$th eigen-operator induced by the reservoir $\mu$, and $\omega^\parallel_{\mu i}$ is the corresponding eigen-frequency.
There are only four allowed transitions for each spin. There are two eigen-operators for the $L$th or the $R$th spin, which can be rewritten as
\begin{alignat}{2}
\nonumber
V^\parallel_{L1}&=\sigma_L^-\otimes m_M^+\otimes\mathbbm{1}_R,\quad &V^\parallel_{L2}&=\sigma_L^-\otimes m_M^-\otimes\mathbbm{1}_R,\\
V^\parallel_{R1}&=\mathbbm{1}_L\otimes m_M^+\otimes\sigma_R^-,\quad &V^\parallel_{R2}&=\mathbbm{1}_L\otimes m_M^-\otimes\sigma_R^-,\label{V_LR}
\end{alignat}
where $m_\mu^+=\vert\uparrow\rangle_\mu\langle\uparrow\vert$ and $m_\mu^-=\vert\downarrow\rangle_\mu\langle\downarrow\vert$. We find that the transitions $V^\parallel_{L1}$ and $V^\parallel_{L2}$ are only caused by the thermal reservoirs $L$ and $M$, independent of the reservoir $R$. Similarly, the transitions of the $R$th spin do not rely on the reservoir $L$. This is caused by the $L$th and the $R$th spins not being directly connected.

The dynamics of the populations can be expressed as
\begin{align}
\nonumber
\dot{\rho}^\parallel_{11}=-\Gamma^{L,\parallel}_{15}-\Gamma^{M,\parallel}_{13}-\Gamma^{R,\parallel}_{12},\quad\dot{\rho}^\parallel_{22}=-\Gamma^{L,\parallel}_{26}-\Gamma^{M,\parallel}_{24}+\Gamma^{R,\parallel}_{12},\\
\nonumber
\dot{\rho}^\parallel_{33}=-\Gamma^{L,\parallel}_{37}+\Gamma^{M,\parallel}_{13}-\Gamma^{R,\parallel}_{34},\quad\dot{\rho}^\parallel_{44}=-\Gamma^{L,\parallel}_{48}+\Gamma^{M,\parallel}_{24}+\Gamma^{R,\parallel}_{34},\\
\nonumber
\dot{\rho}^\parallel_{55}=+\Gamma^{L,\parallel}_{15}-\Gamma^{M,\parallel}_{57}-\Gamma^{R,\parallel}_{56},\quad\dot{\rho}^\parallel_{66}=+\Gamma^{L,\parallel}_{26}-\Gamma^{M,\parallel}_{68}+\Gamma^{R,\parallel}_{56},\\
\dot{\rho}^\parallel_{77}=+\Gamma^{L,\parallel}_{37}+\Gamma^{M,\parallel}_{57}-\Gamma^{R,\parallel}_{78},\quad\dot{\rho}^\parallel_{88}=+\Gamma^{L,\parallel}_{48}+\Gamma^{M,\parallel}_{68}+\Gamma^{R,\parallel}_{78},\label{dynamics}
\end{align} 
where $\Gamma^{\mu,\parallel}_{ij}=2[J_\mu(+\omega_{ij}^{\mu,\parallel})\rho^\parallel_{ii}-J_\mu(-\omega_{ij}^{\mu,\parallel})\rho^\parallel_{jj}]$. %is the net transition rate between the levels $\vert \tilde{i}\rangle$ and $\vert \tilde{j}\rangle$, $\tilde{\rho}_{ii}$ denote the population of the level $\vert \tilde{i}\rangle$. 
According to the evolution equation of the system, we can obtain the unique steady state $\rho^{\parallel, S}$ of the system by the steady-state condition $\vert\dot{\rho}^{\parallel, S}\rangle=\mathcal{M}^\parallel\vert\rho^{\parallel, S}\rangle=0$ and the normalization condition.
Because of its complex expression, the concrete expression of the steady state is not given.

If the middle spin is not in contact with the environment, the dynamics of the populations can be divided into two groups, one for 
\begin{align}
\nonumber
\dot{\tilde{\rho}}^\parallel_{11}=-\tilde{\Gamma}_{15}^{L,\parallel}-\tilde{\Gamma}_{12}^{R,\parallel},\quad\dot{\tilde{\rho}}_{22}^\parallel=-\tilde{\Gamma}_{26}^{L,\parallel}+\tilde{\Gamma}_{12}^{R,\parallel},\\
\dot{\tilde{\rho}}^\parallel_{55}=+\tilde{\Gamma}_{15}^{L,\parallel}-\tilde{\Gamma}_{56}^{R,\parallel},\quad\dot{\tilde{\rho}}_{66}^\parallel=+\tilde{\Gamma}_{26}^{L,\parallel}+\tilde{\Gamma}_{56}^{R,\parallel},\label{rho1256}
\end{align} 
and the other for 
\begin{align}
\nonumber
\dot{\tilde{\rho}}_{33}^\parallel=-\tilde{\Gamma}_{37}^{L,\parallel}-\tilde{\Gamma}_{34}^{R,\parallel},\quad\dot{\tilde{\rho}}_{44}^\parallel=-\tilde{\Gamma}_{48}^{L,\parallel}+\tilde{\Gamma}_{34}^{R,\parallel},\\
\dot{\tilde{\rho}}_{77}^\parallel=+\tilde{\Gamma}_{37}^{L,\parallel}-\tilde{\Gamma}_{78}^{R,\parallel},\quad\dot{\tilde{\rho}}_{88}^\parallel=+\tilde{\Gamma}_{48}^{L,\parallel}+\tilde{\Gamma}_{78}^{R,\parallel},\label{rho3478}
\end{align}
where $\tilde{ }$ denotes the quantity at $\kappa_M=0$. From Eqs. (\ref{rho1256},\ref{rho3478}), the total Hilbert space can be divided into two subspaces. One subspace consists of energy levels $\vert 1^\parallel\rangle$, $\vert 2^\parallel\rangle$, $\vert 5^\parallel\rangle$, and $\vert 6^\parallel\rangle$ named as $S^\parallel_1$, and the other subspace consists of $\vert 3^\parallel\rangle$, $\vert 4^\parallel\rangle$, $\vert 7^\parallel\rangle$, and $\vert 8^\parallel\rangle$) named as $S^\parallel_2$.
In $S^\parallel_l$, $l=1,2$, the evolution equation in matrix form is $\vert\dot{\tilde{\rho}}^\parallel_l(t)\rangle=\tilde{\mathcal{M}}^\parallel_l\vert\tilde{\rho}^\parallel_l(t)\rangle$, where the dissipation coefficient matrix $\tilde{\mathcal{M}}^\parallel_l$ is
\begin{align}
\tilde{\mathcal{M}}^\parallel_l=\sum_{\mu=L,R}\tilde{\mathcal{M}}^{\parallel}_{\mu,l}=\mathrm{M}_L\otimes\mathbbm{1}_2+\mathbbm{1}_2\otimes\mathrm{M}_R
\end{align}
with $\mathrm{M}_\mu=2
\begin{pmatrix}
-J_\mu(-\omega^\parallel_{\mu l})&J_\mu(+\omega^\parallel_{\mu l})\\J_\mu(-\omega^\parallel_{\mu l})&-J_\mu(+\omega^\parallel_{\mu l})
\end{pmatrix}$
and $\vert\tilde{\rho}^\parallel_1(t)\rangle=[\tilde{\rho}^\parallel_{11}(t),\tilde{\rho}^\parallel_{22}(t),\tilde{\rho}^\parallel_{55}(t),\tilde{\rho}^\parallel_{66}(t)]^T$ and $\vert\tilde{\rho}^\parallel_2(t)\rangle=[\tilde{\rho}^\parallel_{33}(t),\tilde{\rho}^\parallel_{44}(t),\tilde{\rho}^\parallel_{77}(t),\tilde{\rho}^\parallel_{88}(t)]^T$.
In the steady state $\tilde{\rho}^{\parallel,S}_l$, the non-zero density matrix elements are
\begin{align}
\nonumber
\tilde{\rho}^{\parallel,S}_{11/33}&=\frac{1}{\tilde{N}_{1/2}^\parallel}J_{L1/2}^{+,\parallel}J_{R1/2}^{+,\parallel},\quad & \tilde{\rho}^{\parallel,S}_{22/44}&=\frac{1}{\tilde{N}_{1/2}^\parallel}J_{L1/2}^{+,\parallel}J_{R1/2}^{-,\parallel},\\
\tilde{\rho}^{\parallel,S}_{55/77}&=\frac{1}{\tilde{N}_{1/2}^\parallel}J_{L1/2}^{-,\parallel}J_{R1/2}^{+,\parallel},\quad & \tilde{\rho}^{\parallel,S}_{66/88}&=\frac{1}{\tilde{N}_{1/2}^\parallel}J_{L1/2}^{-,\parallel}J_{R1/2}^{-,\parallel},\label{rhoSLF}
\end{align}
where $J_{\mu l}^{\pm,\parallel}=J_\mu(\pm\omega^\parallel_{\mu l})$ for the sake of brevity of the expression and $\tilde{N}_l^\parallel=(J_{Ll}^{+,\parallel}+J_{Ll}^{-,\parallel})(J_{Rl}^{+,\parallel}+J_{Rl}^{-,\parallel})$ is the normalization coefficient.

In this case, the steady-state heat currents are
\begin{align}
\nonumber
\dot{\tilde{Q}}^\parallel_L&=-\tilde{p}\omega^\parallel_{L1}(\tilde{\Gamma}^{L,\parallel}_{15}+\tilde{\Gamma}_{26}^{L,\parallel})-(1-\tilde{p})\omega_{L2}^\parallel(\tilde{\Gamma}_{37}^{L,\parallel}+\tilde{\Gamma}_{48}^{L,\parallel})=0,\\
\dot{\tilde{Q}}^\parallel_R&=-\tilde{p}\omega^\parallel_{R1}(\tilde{\Gamma}^{R,\parallel}_{12}+\tilde{\Gamma}_{56}^{R,\parallel})-(1-\tilde{p})\omega_{R2}^\parallel(\tilde{\Gamma}_{34}^{R,\parallel}+\tilde{\Gamma}_{78}^{R,\parallel})=0.\label{A10}
\end{align}
Substituting the steady state Eq. (\ref{rhoSLF}) into the net transition rate $\tilde{\Gamma}_{ij}^{\mu,\parallel}$, it is easy to obtain $\tilde{\Gamma}_{ij}^{\mu,\parallel}=0$. Thus, when only the LF is present and the middle spin is not in contact with the environment, the steady-state heat currents of the system all vanish, which is independent of the fractions of two subspaces.

One can find that in the case $\kappa_M\neq 0$, the system can also block the heat current when the amplitude of the magnetic field is precisely equal to the coupling strength between the spins, i.e., $B_\mu=J_{\mu\nu}\equiv B$. In this case, the energy levels $\vert 3^\parallel\rangle$, $\vert 4^\parallel\rangle$, $\vert 7^\parallel\rangle$, and $\vert 8^\parallel\rangle$ are degenerated. The eigen-frequencies of the system are $\check{\omega}_{51}^{L,\parallel}=\check{\omega}_{62}^{L,\parallel}=\check{\omega}_{21}^{R,\parallel}=\check{\omega}_{65}^{R,\parallel}=2B$, $\check{\omega}_{73}^{L,\parallel}=\check{\omega}_{84}^{L,\parallel}=\check{\omega}_{43}^{R,\parallel}=\check{\omega}_{87}^{R,\parallel}=0$, $\check{\omega}_{31}^{M,\parallel}=3B$, and $\check{\omega}^{M,\parallel}_{42}=\check{\omega}^{M,\parallel}_{75}=\check{\omega}^{M,\parallel}_{68}=B$, where $\check{ }$ represents the quantity in the condition $B_\mu=J_{\mu\nu}$. Therefore, the net transition rates are $\check{\Gamma}^{L,\parallel}_{37}=\check{\Gamma}^{L,\parallel}_{48}=\check{\Gamma}^{R,\parallel}_{34}=\check{\Gamma}^{R,\parallel}_{78}=0$ in the dynamics Eq. (\ref{dynamics}).
Following the same steps, we obtain that the steady-state density matrix as $\check{\rho}^{\parallel, S}=\sum_{i=1}^8\check{\rho}^{\parallel, S}_{ii}\vert i^\parallel\rangle\langle i^\parallel\vert=\frac{1}{\check{N}}\sum_{i=1}^8\check{\rho}_{i}^{\parallel, S}\vert i^\parallel\rangle\langle i^\parallel\vert$, which $\check{N}=\sum_{i=1}^8\check{\rho}_i^{\parallel, S}$ is the normalized coefficient and the non-normalized diagonal elements are
\begin{align}
\nonumber
\check{\rho}_{1}^{\parallel,S}=J_L(+2B)J_M(+3B)J_M(-B)J_M(+B)J_R(+2B),\\
\nonumber
\check{\rho}_{2}^{\parallel,S}=J_L(+2B)J_M(+3B)J_M(-B)J_M(+B)J_R(-2B),\\
\nonumber
\check{\rho}_{3}^{\parallel,S}=J_L(+2B)J_M(-3B)J_M(-B)J_M(+B)J_R(+2B),\\
\nonumber
\check{\rho}_{4}^{\parallel,S}=J_L(+2B)J_M(+3B)J_M(-B)J_M(-B)J_R(-2B),\\
\nonumber
\check{\rho}_{5}^{\parallel,S}=J_L(-2B)J_M(+3B)J_M(-B)J_M(+B)J_R(+2B),\\
\nonumber
\check{\rho}_{6}^{\parallel,S}=J_L(-2B)J_M(+3B)J_M(-B)J_M(+B)J_R(-2B),\\
\nonumber
\check{\rho}_{7}^{\parallel,S}=J_L(-2B)J_M(+3B)J_M(-B)J_M(-B)J_R(+2B),\\
\check{\rho}_{8}^{\parallel,S}=J_L(-2B)J_M(+3B)J_M(+B)J_M(+B)J_R(-2B).
\end{align}
According to the definition of the heat current Eq. (\ref{Qdefinationde}), we get the steady-state heat current as
\begin{align}
\dot{\check{Q}}^\parallel_L=\dot{\check{Q}}^\parallel_M=\dot{\check{Q}}^\parallel_R=0.
\end{align}

\section{Detailed derivation of the steady state in TF}
\label{AppendixB}
If the system is immersed in the TF, the Hamiltonian of the system is
\begin{align}
H_S^\perp=\frac{1}{2}(\sum_\mu B_\mu\sigma^x_\mu+\sum_{\mu\neq \nu}J_{\mu\nu}\sigma_\mu^z\sigma_\nu^z),\quad\mu,\nu=L,M,R.\label{HS3TF}
\end{align}
The system is diagonalized as $H_S^\perp=\sum_i\omega^\perp_i\vert i^\perp\rangle\langle i^\perp\vert$, where $\omega^\perp_i$ and $\vert i^\perp\rangle$ are the eigenvalue and the corresponding eigenstate. Since the amplitude of the magnetic field of each spin $B_\mu$ has no qualitative effect on the characteristic of the system, we consider here only the case of $B_\mu=B$. The eigenvalues are
\begin{align}
\nonumber
\omega^\perp_1&=-\frac{1}{4\sqrt{6}}(\sqrt{m_3^{-+}}+\sqrt{m_2^-}), &\omega^\perp_2&=\frac{1}{4\sqrt{6}}(\sqrt{m_3^{-+}}-\sqrt{m_2^-}),\\
\nonumber
\omega^\perp_3&=-\frac{1}{4\sqrt{6}}(\sqrt{m_3^{--}}-\sqrt{m_2^-}), &\omega^\perp_4&=\frac{1}{4\sqrt{6}}(\sqrt{m_3^{--}}+\sqrt{m_2^-}),\\
\nonumber
\omega^\perp_5&=-\frac{1}{4\sqrt{6}}(\sqrt{m_3^{++}}+\sqrt{m_2^+}), &\omega^\perp_6&=\frac{1}{4\sqrt{6}}(\sqrt{m_3^{++}}-\sqrt{m_2^+}),\\
\omega^\perp_7&=-\frac{1}{4\sqrt{6}}(\sqrt{m_3^{+-}}-\sqrt{m_2^+}), &\omega^\perp_8&=\frac{1}{4\sqrt{6}}(\sqrt{m_3^{+-}}+\sqrt{m_2^+}),\label{eigenvalues}
\end{align}
and the parameters in the above equation are
\begin{align}
\nonumber
m^{\pm\mp}_3&=-12c-m_2^\pm\mp\frac{12\sqrt{6}d^\pm}{\sqrt{m_2^\pm}},\\
\nonumber
m_2^\pm&=m_1^\pm+\frac{4(c^2+12e)}{m_1^\pm}-4c,\\
\nonumber
m_1^\pm&=\{4[2c^3+27{d^\pm}^2-72ce\\
\nonumber
&+\sqrt{{(2c^3+27{d^\pm}^2-72ce)}^2-4(c^2+12e)^3}] \}^{1/3},\\
\nonumber
c&=-6B^2-2J_{LM}^2-2J_{MR}^2,\\
\nonumber
d^\pm&=\pm8B^3,\\
e&=-3B^4+2B^2(J_{LM}^2+J_{MR}^2)+(J_{LM}^2-J_{MR}^2)^2.
\end{align}
The eigenstates are $\vert i^\perp\rangle=\sum_{j=1}^8\Lambda^\perp(i,j)\vert\tilde{j}\rangle$, where $\Lambda^\perp(i,j)$ is the matrix element of $ith$ row and $jth$ column of coefficient matrix $\Lambda^\perp$ represented as
\begin{equation}
\Lambda^\perp=\begin{pmatrix}
\Lambda_{11} & \Lambda_{12} & \Lambda_{13} & \Lambda_{14} & \Lambda_{14} & \Lambda_{13} & \Lambda_{12} & \Lambda_{11}\\
\Lambda_{21} & \Lambda_{22} & \Lambda_{23} & \Lambda_{24} & \Lambda_{24} & \Lambda_{23} & \Lambda_{22} & \Lambda_{21}\\
\Lambda_{31} & \Lambda_{32} & \Lambda_{33} & \Lambda_{34} & \Lambda_{34} & \Lambda_{33} & \Lambda_{32} & \Lambda_{31}\\
\Lambda_{41} & \Lambda_{42} & \Lambda_{43} & \Lambda_{44} & \Lambda_{44} & \Lambda_{43} & \Lambda_{42} & \Lambda_{41}\\
-\Lambda_{51} & \Lambda_{52} & -\Lambda_{53} & -\Lambda_{54} & \Lambda_{54} & \Lambda_{53} & -\Lambda_{52} & \Lambda_{51}\\
-\Lambda_{61} & \Lambda_{62} & -\Lambda_{63} & -\Lambda_{64} & \Lambda_{64} & \Lambda_{63} & -\Lambda_{62} & \Lambda_{61}\\
-\Lambda_{71} & \Lambda_{72} & -\Lambda_{73} & -\Lambda_{74} & \Lambda_{74} & \Lambda_{73} & -\Lambda_{72} & \Lambda_{71}\\
-\Lambda_{81} & \Lambda_{82} & -\Lambda_{83} & -\Lambda_{84} & \Lambda_{84} & \Lambda_{83} & -\Lambda_{82} & \Lambda_{81}
\end{pmatrix}.\label{LambdaHS3}
\end{equation}
Defining some parameters is necessary to represent the matrix elements of the above matrix in concrete terms. Firstly, for $i=1,\cdot\cdot\cdot,4$, there are
\begin{align}
\nonumber
\lambda_i^{\pm\mp}&=B\pm J_{LM}\mp J_{MR}+\omega_i^\perp,\\
\nonumber
\mathit{\Lambda}_{i2}&=\frac{\lambda_i^{--}}{\lambda_i^{-+}},\mathit{\Lambda}_{i3}=\frac{\breve{\mathit{\Lambda}}_{i3}}{B\omega^\perp_i\lambda_i^{-+}\lambda_i^{+-}},\mathit{\Lambda}_{i4}=\frac{\lambda_i^{--}}{\lambda_i^{+-}},\\
\nonumber
\breve{\mathit{\Lambda}}_{i3}&=3B^4-(J_{LM}^2-J_{MR}^2)^2-2B^2(J_{LM}^2+J_{MR}^2)\\
\nonumber
&+\{6B^3+[B^2+(J_{LM}-J_{MR})^2](J_{LM}+J_{MR})\}\omega^\perp_i\\
&+[3B^2+(J_{LM}+J_{MR})^2]{\omega^\perp_i}^2-(J_{LM}+J_{MR}){\omega^\perp_i}^3,\label{para1}
\end{align}
and for $i=5,\cdot\cdot\cdot,8$, there are
\begin{align}
\nonumber
\lambda_i^{\pm\mp}&=B\pm J_{LM}\mp J_{MR}-\omega^\perp_i,\\
\nonumber
\mathit{\Lambda}_{i2}&=\frac{\lambda_i^{++}}{\lambda_i^{+-}},\mathit{\Lambda}_{i3}=\frac{\breve{\mathit{\Lambda}}_{i3}}{B\omega^\perp_i\lambda_i^{-+}\lambda_i^{+-}},\mathit{\Lambda}_{i4}=\frac{\lambda_i^{++}}{\lambda_i^{-+}},\\
\nonumber
\breve{\mathit{\Lambda}}_{i3}&=3B^4-(J_{LM}^2-J_{MR}^2)^2-2B^2(J_{LM}^2+J_{MR}^2)\\
\nonumber
&+\{-6B^3+[B^2+(J_{LM}-J_{MR})^2](J_{LM}+J_{MR})\}\omega^\perp_i\\
&+[3B^2+(J_{LM}+J_{MR})^2]{\omega^\perp_i}^2-(J_{LM}+J_{MR}){\omega^\perp_i}^3.\label{para2}
\end{align}
Using the parameters of Eqs. (\ref{para1}, \ref{para2}), the unnormalized coefficient matrix $\mathit{\Lambda}^\perp$, i.e., $\langle\breve{i}\vert\breve{i}\rangle\neq 1$ with $\vert\breve{i}\rangle=\sum_{j=1}^8\mathit{\Lambda}^\perp(i,j)\vert\tilde{j}\rangle$, as
\begin{equation}
\mathit{\Lambda}^\perp=\begin{pmatrix}
1 & \mathit{\Lambda}_{12} & \mathit{\Lambda}_{13} & \mathit{\Lambda}_{14} & \mathit{\Lambda}_{14} & \mathit{\Lambda}_{13} & \mathit{\Lambda}_{12} & 1\\
1 & \mathit{\Lambda}_{22} & \mathit{\Lambda}_{23} & \mathit{\Lambda}_{24} & \mathit{\Lambda}_{24} & \mathit{\Lambda}_{23} & \mathit{\Lambda}_{22} & 1\\
1 & \mathit{\Lambda}_{32} & \mathit{\Lambda}_{33} & \mathit{\Lambda}_{34} & \mathit{\Lambda}_{34} & \mathit{\Lambda}_{33} & \mathit{\Lambda}_{32} & 1\\
1 & \mathit{\Lambda}_{42} & \mathit{\Lambda}_{43} & \mathit{\Lambda}_{44} & \mathit{\Lambda}_{44} & \mathit{\Lambda}_{43} & \mathit{\Lambda}_{42} & 1\\
-1 & \mathit{\Lambda}_{52} & -\mathit{\Lambda}_{53} & -\mathit{\Lambda}_{54} & \mathit{\Lambda}_{54} & \mathit{\Lambda}_{53} & -\mathit{\Lambda}_{52} & 1\\
-1 & \mathit{\Lambda}_{62} & -\mathit{\Lambda}_{63} & -\mathit{\Lambda}_{64} & \mathit{\Lambda}_{64} & \mathit{\Lambda}_{63} & -\mathit{\Lambda}_{62} & 1\\
-1 & \mathit{\Lambda}_{72} & -\mathit{\Lambda}_{73} & -\mathit{\Lambda}_{74} & \mathit{\Lambda}_{74} & \mathit{\Lambda}_{73} & -\mathit{\Lambda}_{72} & 1\\
-1 & \mathit{\Lambda}_{82} & -\mathit{\Lambda}_{83} & -\mathit{\Lambda}_{84} & \mathit{\Lambda}_{84} & \mathit{\Lambda}_{83} & -\mathit{\Lambda}_{82} & 1\\
\end{pmatrix}.
\end{equation}
Then, based on the normalization condition $\langle i^\perp\vert i^\perp\rangle =1$, we can obtain Eq. (\ref{LambdaHS3}).

The eigen-operator and the corresponding eigenfrequency induced by the thermal reservoir $\mu$ are $V^{\mu,\perp}_{ij}=\Lambda^{\mu,\perp}_{ij}\vert i^\perp\rangle\langle j^\perp\vert$ and $\omega^{\mu,\perp}_{ij}$, respectively, where the coefficients are
\begin{align}
\nonumber
\Lambda^{L,\perp}_{ij}=2[+\sum_{k=1,4}\Lambda^\perp_{ik}\Lambda^\perp_{j5-k}+\sum_{k=2,3}\Lambda^\perp_{ik}\Lambda^\perp_{j5-k}], \\
\nonumber
\Lambda^{M,\perp}_{ij}=2[+\sum_{k=1,3}\Lambda^\perp_{ik}\Lambda^\perp_{j4-k}+\sum_{k=2,4}\Lambda^\perp_{ik}\Lambda^\perp_{j6-k}], \\
\Lambda^{R,\perp}_{ij}=2[+\sum_{k=1,2}\Lambda^\perp_{ik}\Lambda^\perp_{j3-k}+\sum_{k=3,4}\Lambda^\perp_{ik}\Lambda^\perp_{j7-k}], \label{1234}
\end{align}
for $i\in[1,3]$, $j\in[i+1,4]$, and
\begin{align}
\nonumber
\Lambda^{L,\perp}_{ij}=2[-\sum_{k=1,4}\Lambda^\perp_{ik}\Lambda^\perp_{j5-k}+\sum_{k=2,3}\Lambda^\perp_{ik}\Lambda^\perp_{j5-k}],\\
\nonumber
\Lambda^{M,\perp}_{ij}=2[-\sum_{k=1,3}\Lambda^\perp_{ik}\Lambda^\perp_{j4-k}+\sum_{k=2,4}\Lambda^\perp_{ik}\Lambda^\perp_{j6-k}],\\
\Lambda^{R,\perp}_{ij}=2[-\sum_{k=1,2}\Lambda^\perp_{ik}\Lambda^\perp_{j3-k}+\sum_{k=3,4}\Lambda^\perp_{ik}\Lambda^\perp_{j7-k}],\label{5678}
\end{align}
for $i\in[5,7]$, $j\in[i+1,8]$.
Except for the 12 transitions mentioned above for each spin, other transitions are not allowed.
The transition $\vert i^\perp\rangle\leftrightarrow\vert j^\perp\rangle$ can be induced by either thermal reservoir, although the probability of transition $\Lambda_{ij}^{\mu,\perp}$ is different for each thermal reservoir.

Since only the steady-state property of the system is considered, the dynamics of the populations, which are non-zero at steady state, can be represented as two separate sets of groups based on Eqs. (\ref{1234},\ref{5678},\ref{dynamic_total}); one group is
\begin{align}
\nonumber
\dot{\rho}^\perp_{11}&=\sum_\mu(-\Gamma_{12}^{\mu,\perp}-\Gamma_{13}^{\mu,\perp}-\Gamma_{14}^{\mu,\perp}),\\
\nonumber
\dot{\rho}^\perp_{22}&=\sum_\mu(+\Gamma_{12}^{\mu,\perp}-\Gamma_{23}^{\mu,\perp}-\Gamma_{24}^{\mu,\perp}),\\
\nonumber
\dot{\rho}^\perp_{33}&=\sum_\mu(+\Gamma_{13}^{\mu,\perp}+\Gamma_{23}^{\mu,\perp}-\Gamma_{34}^{\mu,\perp}),\\
\dot{\rho}^\perp_{44}&=\sum_\mu(+\Gamma_{14}^{\mu,\perp}+\Gamma_{24}^{\mu,\perp}+\Gamma_{34}^{\mu,\perp}),\label{rho1234}
\end{align}
and the other group is
\begin{align}
\nonumber
\dot{\rho}^\perp_{55}&=\sum_\mu(-\Gamma_{56}^{\mu,\perp}-\Gamma_{57}^{\mu,\perp}-\Gamma_{58}^{\mu,\perp}),\\
\nonumber
\dot{\rho}^\perp_{66}&=\sum_\mu(+\Gamma_{56}^{\mu,\perp}-\Gamma_{67}^{\mu,\perp}-\Gamma_{68}^{\mu,\perp}),\\
\nonumber
\dot{\rho}^\perp_{77}&=\sum_\mu(+\Gamma_{57}^{\mu,\perp}+\Gamma_{67}^{\mu,\perp}-\Gamma_{78}^{\mu,\perp}),\\
\dot{\rho}^\perp_{88}&=\sum_\mu(+\Gamma_{58}^{\mu,\perp}+\Gamma_{68}^{\mu,\perp}+\Gamma_{78}^{\mu,\perp}).\label{rho5678}
\end{align}

The net transition rate is $\Gamma^{\mu,\perp}_{ij}=2(\Lambda_{ij}^{\mu,\perp})^2[J_\mu(+\omega_{ij}^{\mu,\perp})\rho^\perp_{ii}-J_\mu(-\omega_{ij}^{\mu,\perp})\rho^\perp_{jj}]$. Obviously, the system is automatically decoupled into two independent subspaces $S^\perp_1$ and $S^\perp_2$. Energy levels $\vert 1^\perp\rangle$, $\vert  2^\perp\rangle$, $\vert 3^\perp\rangle$, and $\vert 4^\perp\rangle$ constitute the subspace $S_1^\perp$, and the other four energy levels constitute the subspace $S_2^\perp$.
In $S^\perp_1$, the dynamic is $\vert\dot{\rho}^\perp_1(t)\rangle=\mathcal{M}^\perp_1\vert\rho^\perp_1(t)\rangle$, where $\vert\rho^\perp_1(t)\rangle=[\rho^\perp_{11}(t),\rho^\perp_{22}(t),\rho^\perp_{33}(t),\rho^\perp_{44}(t)]^T$. Thus, the steady-state dynamics in matrix form can be expressed as
\begin{widetext}
\begin{align}
\begin{pmatrix}
-M_{12}^+-M_{13}^+-M_{14}^+ & M_{12}^- & M_{13}^- & M_{14}^-\\
M_{12}^+ & -M_{12}^--M_{23}^+-M_{24}^+ & M_{23}^- & M_{24}^-\\
M_{13}^+ & M_{23}^+ & -M_{13}^--M_{23}^--M_{34}^+ & M_{34}^-\\
M_{14}^+ & M_{24}^+ & M_{34}^+ & -M_{14}^--M_{24}^--M_{34}^-
\end{pmatrix}
\begin{pmatrix}
\rho_{11}^{\perp,S}\\\rho_{22}^{\perp,S}\\\rho_{33}^{\perp,S}\\\rho_{44}^{\perp,S}
\end{pmatrix}
=0,
\end{align}
where $M_{ij}^\pm=\sum_\mu M_{ij}^{\mu,\pm}=\sum_\mu 2(\Lambda_{ij}^{\mu,\perp})^2J_\mu(\pm\omega_{ij}^{\mu,\perp})$.
The steady state in $S^\perp_1$ is $\rho_1^{\perp,S}=\sum_{i=1}^4\rho_{ii}^{\perp,S}\vert i^\perp\rangle\langle i^\perp\vert$, where $\rho_{ii}^{\perp,S}=\frac{\breve{\rho}_{ii}}{N_1}$ is the normalized population with the normalized coefficient $N_1=\sqrt{\sum_{i=1}^4{\breve{\rho}_{ii}}^2}$ and
\begin{align}
\nonumber
\breve{\rho}_{11}&=M_{12}^-M_{13}^-M_{14}^-+M_{12}^-M_{13}^-M_{24}^-+M_{12}^-M_{13}^-M_{34}^-+M_{12}^-M_{14}^-M_{23}^-+M_{12}^-M_{14}^-M_{34}^++M_{12}^-M_{23}^-M_{24}^-+M_{12}^-M_{23}^-M_{34}^-+M_{12}^-M_{24}^-M_{34}^+\\
\nonumber
&+M_{13}^-M_{14}^-M_{23}^++M_{13}^-M_{14}^-M_{24}^++M_{13}^-M_{23}^+M_{24}^-+M_{13}^-M_{23}^+M_{34}^-+M_{13}^-M_{24}^+M_{34}^-+M_{14}^-M_{23}^-M_{24}^++M_{14}^-M_{23}^+M_{34}^++M_{14}^-M_{24}^+M_{34}^+,\\
\nonumber
\breve{\rho}_{22}&=M_{12}^+M_{13}^-M_{14}^-+M_{12}^+M_{13}^-M_{24}^-+M_{12}^+M_{13}^-M_{34}^-+M_{12}^+M_{14}^-M_{23}^-+M_{12}^+M_{14}^-M_{34}^++M_{12}^+M_{23}^-M_{24}^-+M_{12}^+M_{23}^-M_{34}^-+M_{12}^+M_{24}^-M_{34}^+\\
\nonumber
&+M_{13}^+M_{14}^-M_{23}^-+M_{13}^-M_{14}^+M_{24}^-+M_{13}^+M_{23}^-M_{24}^-+M_{13}^+M_{23}^-M_{34}^-+M_{13}^+M_{24}^-M_{34}^++M_{14}^+M_{23}^-M_{24}^-+M_{14}^+M_{23}^-M_{34}^-+M_{14}^+M_{24}^-M_{34}^+,\\
\nonumber
\breve{\rho}_{33}&=M_{12}^-M_{13}^+M_{14}^-+M_{12}^-M_{13}^+M_{24}^-+M_{12}^-M_{13}^+M_{34}^-+M_{12}^+M_{14}^-M_{23}^++M_{12}^-M_{14}^+M_{34}^-+M_{12}^+M_{23}^+M_{24}^-+M_{12}^+M_{23}^+M_{34}^-+M_{12}^+M_{24}^+M_{34}^-\\
\nonumber
&+M_{13}^+M_{14}^-M_{23}^++M_{13}^+M_{14}^-M_{24}^++M_{13}^+M_{23}^+M_{24}^-+M_{13}^+M_{23}^+M_{34}^-+M_{13}^+M_{24}^+M_{34}^-+M_{14}^+M_{23}^+M_{24}^-+M_{14}^+M_{23}^+M_{34}^-+M_{14}^+M_{24}^+M_{34}^-,\\
\nonumber
\breve{\rho}_{44}&=M_{12}^-M_{13}^-M_{14}^++M_{12}^+M_{13}^-M_{24}^++M_{12}^-M_{13}^+M_{34}^++M_{12}^-M_{14}^+M_{23}^-+M_{12}^-M_{14}^+M_{34}^++M_{12}^+M_{23}^-M_{24}^++M_{12}^+M_{23}^+M_{34}^++M_{12}^+M_{24}^+M_{34}^+\\
&+M_{13}^-M_{14}^+M_{23}^++M_{13}^-M_{14}^+M_{24}^++M_{13}^+M_{23}^-M_{24}^++M_{13}^+M_{23}^+M_{34}^++M_{13}^+M_{24}^+M_{34}^++M_{14}^+M_{23}^-M_{24}^++M_{14}^+M_{23}^+M_{34}^++M_{14}^+M_{24}^+M_{34}^+.\label{steady_TF}
\end{align}
\end{widetext}

\begin{figure*}
	\centering
		\includegraphics[width=18cm]{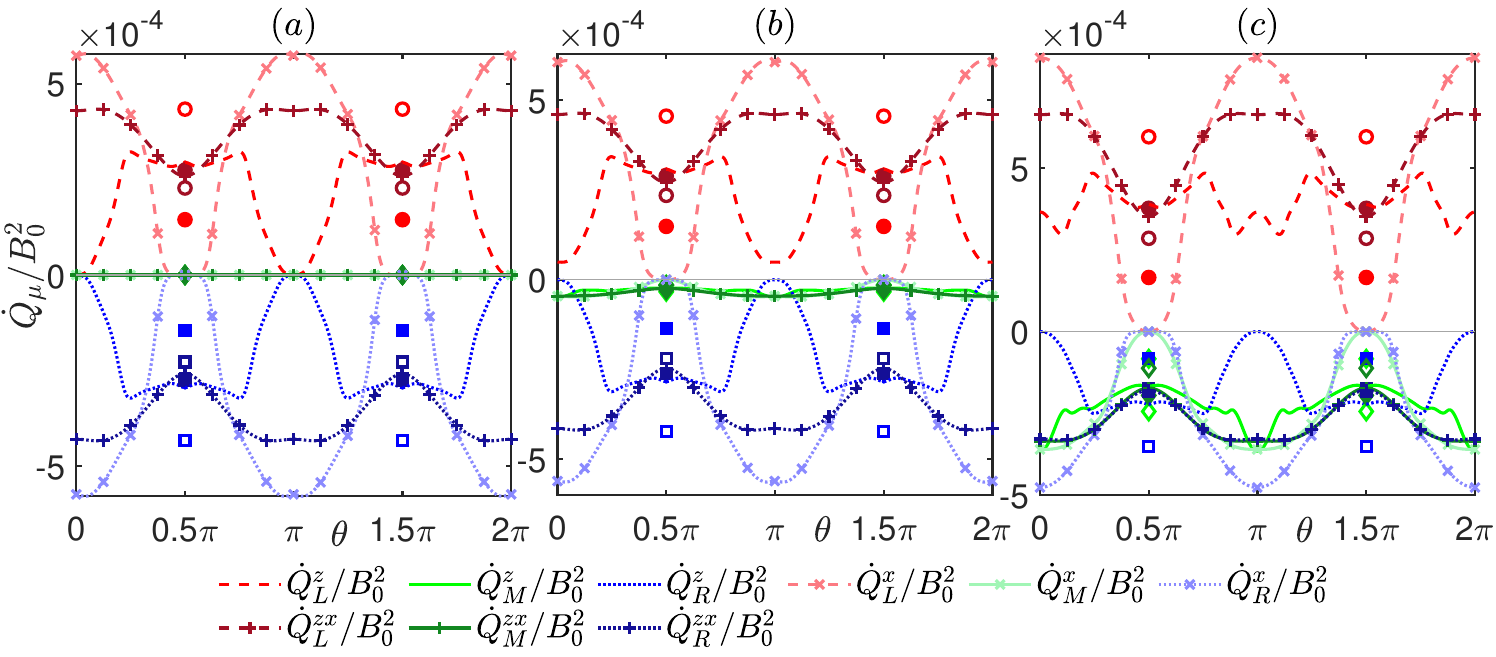}
	\caption{The process of modulating heat currents by adjusting the direction of the magnetic field under different types of couplings between spins. The unmarked, cross, and plus lines correspond to $\sigma_\mu^z\sigma_\nu^z$, $\sigma_\mu^x\sigma_\nu^x$, and $\sigma_\mu^z\sigma_\nu^z+\sigma_\mu^x\sigma_\nu^x$ coupling modes between spins, respectively. Dashed, solid and dotted lines correspond to $\dot{Q}_L$, $\dot{Q}_M$, and $\dot{Q}_R$, respectively. Solid (or hollow) circles, diamonds, and squares correspond to $\dot{Q}_L$, $\dot{Q}_M$, and $\dot{Q}_R$ for $p=0$ (or $p=1$) when $\theta=0.5\pi$. Note that when the coupling is $\sigma_\mu^x\sigma_\nu^x$, the heat current in the presence of the transverse field is 0, which is not represented in the figure. Here, $B_0=1$, $B_L=B_M=B_R=1.5B_0$, $J_{LM}=0.8B_0$, $J_{MR}=1.2B_0$, $T_L=2B_0$, $T_M=0.02B_0$, $T_R=0.2B_0$, $\kappa_L=\kappa_R=0.001B_0$, $\kappa_M=0$, $0.1\kappa_L$, and $\kappa_L$ in (a), (b), and (c).}
\label{Q_zx}
\end{figure*}

Replacing the four energy levels $\vert 1^\perp\rangle$, $\vert 2^\perp\rangle$, $\vert 3^\perp\rangle$, and $\vert 4^\perp\rangle$ of $S_1^\perp$ with the other four energy levels $\vert 5^\perp\rangle$, $\vert 6^\perp\rangle$, $\vert 7^\perp\rangle$, and $\vert 8^\perp\rangle$, we can obtain the non-zero density matrix elements $\rho_{55}^{\perp,S}$, $\rho_{66}^{\perp,S}$, $\rho_{77}^{\perp,S}$, and $\rho_{88}^{\perp,S}$ in the steady state of the subspace $S^\perp_2$, which is not repeated here.

\section{Modulation of heat current for coupling in different directions}
\label{AppendixC}
In this appendix, we study the effects of coupling interactions in different directions on heat current. The most general XYZ Heisenberg interaction between the spins is
\begin{align}
H_{SI}=\sum_{(\mu,\nu)}\frac{J_{\mu\nu}}{2}(g^x\sigma_\mu^x\sigma_\nu^x+g^y\sigma_\mu^y\sigma_\nu^y+g^z\sigma_\mu^z\sigma_\nu^z),
\end{align}
for $(\mu,\nu)\in\{(L,M),(M,R)\}$, where $g^x,g^y,g^z\in[0,1]$ represents the coupled anisotropy of the two nearest-neighbor spins in the $x,y,z$ direction. We use $(g^xg^yg^z)$ to indicate the case of different anisotropy, $g^x=0$ and $g^x\neq 0$ are denoted by the symbols $\circ$ and $\bullet$, respectively.

\begin{figure*}
	\centering
		\includegraphics[width=18cm]{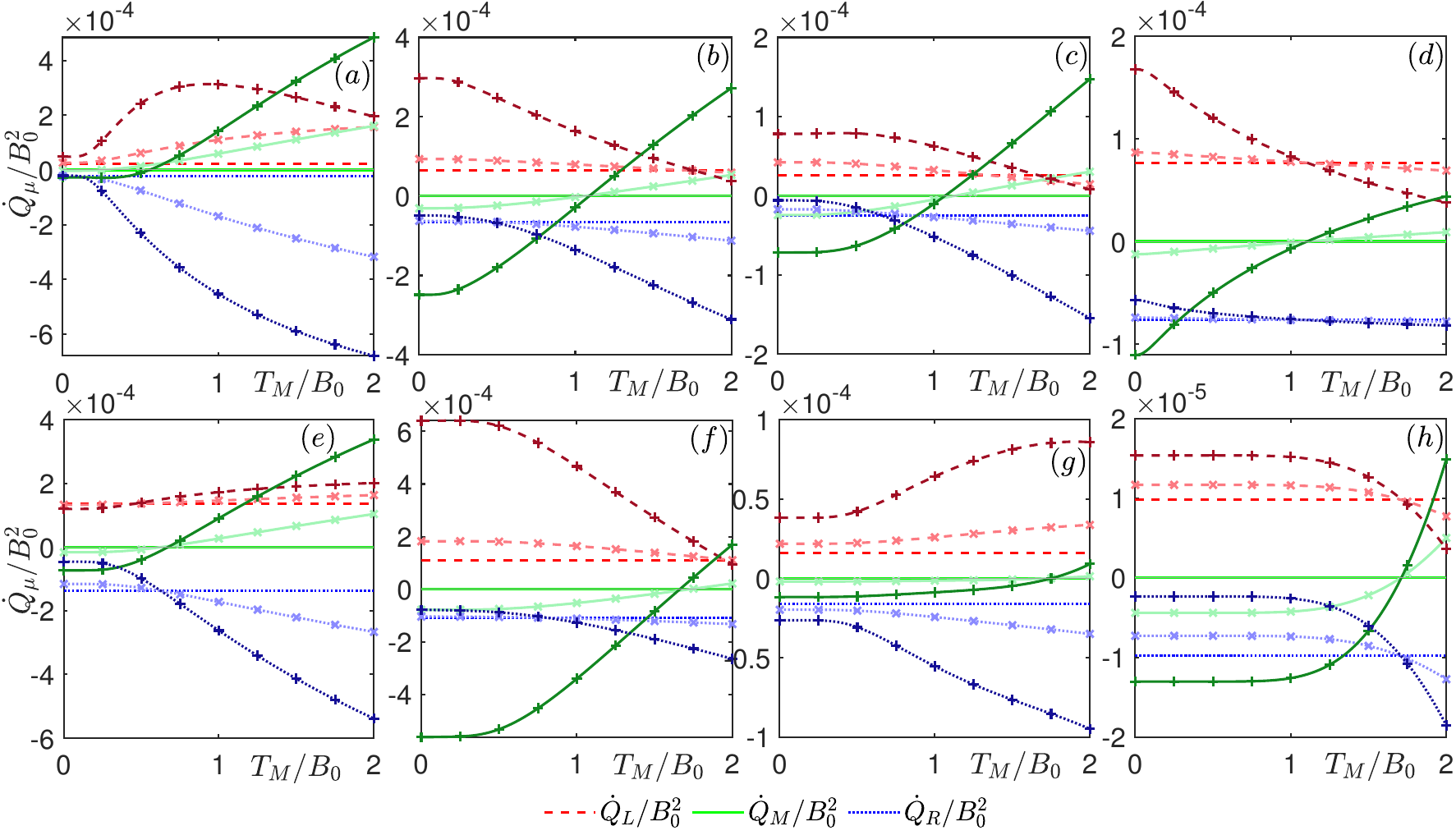}
	\caption{Steady-state heat currents vary with the temperature of the middle heat reservoir $T_M$. The selection of parameters and the meaning of the line styles in this figure are consistent with those in Fig. \ref{Q_theta} and the direction of the magnetic field is fixed $\theta=0.1\pi$. }
\label{Q_TM}
\end{figure*}

When only the longitudinal field is present, the steady state of the system is unique and the steady-state heat current is not blocked except for the case $(\circ\circ\bullet)$ with $\kappa_M=0$.
In the transverse field case, we find that the eigenstates in any coupling type can be expressed as Eq. (\ref{LambdaHS3}) except for two cases, which are $(\bullet\circ\circ)$ and isotropic $(\circ\bullet\bullet)$. Therefore, the stable system in these cases can also be represented as a direct sum space of two subspaces. Table \ref{table} gives the number of steady states of the system when only transverse field exists with different coupling types $(g^xg^yg^z)$, where \textit{isotropy} and \textit{anisotropy} represent the same and different coupling in different directions when the nearest-neighbor spins are coupled in more than one direction. 
For case $(\bullet\circ\circ)$, the eigenstates of the system in the present of transverse field are $\vert +++\rangle$, $\vert ++-\rangle$, $\vert +-+\rangle$, $\vert +--\rangle$, $\vert -++\rangle$, $\vert -+-\rangle$, $\vert --+\rangle$, and $\vert ---\rangle$, where $\vert\pm\rangle_\mu=\frac{1}{\sqrt{2}}(\vert\uparrow\rangle_\mu\pm\vert\downarrow\rangle_\mu)$ are the eigenstate of the operator $\sigma_\mu^x$. For spin-Boson coupling $\sum_k\sigma^x_\mu(a_{\mu k}+a^\dagger_{\mu k})$, there is no transition between any two eigenstates since the eigen-operators are all 0, and the space formed by any eigenstate is an invariant subspace. Therefore, when the coupling between the nearest-neighbor spins is $\sigma^x_\mu\sigma^x_\nu$, the heat current is also disappeared. The physics behind the vanished heat current can be understood as follows.
For the general spin-Boson model $H_{S-B}=\frac{1}{2}\Delta\sigma^x+\sum_{k,\mu}\omega_{k,\mu}a_{k,\mu}^\dagger a_{k,\mu}+\sum_{k,\mu}(g_{k,\mu}^z\sigma^z+g_{k,\mu}^x\sigma^x)(a_{k,\mu}^\dagger+a_{k,\mu})$, $g_{k,\mu}^z$ and $g_{k,\mu}^x$ describe the coupling strength of dissipative and dephasing interactions between the qubit and the corresponding reservoir, respectively. For $g_{k,\mu}^z=0$, the dephasing interaction results in pure decoherence, no energy loss (i.e., heat current is zero) is a trivial result \cite{Cattaneo_2019}.
For the isotropic $(\circ\bullet\bullet)$, i.e., $g^y=g^z\neq 0$, two of the eight eigenstates of the system become $\vert +++\rangle$ and $\vert ---\rangle$, and the other six eigenstates can still be divided into two subspaces, so there are four subspaces. 
The heat current modulation processes under three coupling types, which are $\sigma_\mu^z\sigma_\nu^z$, $\sigma_\mu^x\sigma_\nu^x$, and $\sigma_\mu^x\sigma_\nu^x+\sigma_\mu^z\sigma_\nu^z$, are shown in Fig. \ref{Q_zx} with the unmarked, cross, and plus lines.
Regardless of the coupling type, the heat currents periodically vary with the direction of the magnetic field. But only for coupling $\sigma^z_\mu\sigma^z_\nu$, in the case $\kappa_M=0$, the heat current can be perfectly modulated in the dissipative dynamics.
\begin{table}[h!]
  \centering
  \caption{The number of steady states resulting from different coupling $(g^x g^y g^z)$ between the spins in the present of TF.}
  \begin{tabular}{c|ccccccc}
    \hline\hline
%   $(g^x g^y g^z)$ & $(0 0 1)$ & $(0 1 0)$ & $(1 0 0)$ & $(1 0 1)$ & $(0 1 1)$ & $(1 1 0)$ & $(1 1 1)$
   $(g^x g^y g^z)$ & $(\circ\circ\bullet)$ & $(\circ\bullet\circ)$ & $(\bullet\circ\circ)$ & $(\bullet\circ\bullet)$ & $(\circ\bullet\bullet)$ & $(\bullet\bullet\circ)$ & $(\bullet\bullet\bullet)$
     \rule{0pt}{2.6ex}\rule[-1.2ex]{0pt}{0pt}
    \\ \hline
    isotropic  & 2 & 2 & 8 & 2 & 4 & 2 & 2 \\
    \hline
    anisotropic   &   &   &   & 2 & 2 & 2 & 2 \\ \hline\hline
  \end{tabular}\label{table}
\end{table}

\section{Variation of Steady-state heat current with $T_M$}
\label{AppendixD}
Fig. \ref{Q_TM} gives the variations of the steady-state heat currents from the three heat reservoirs to the system with the temperature of the middle reservoir. The competition between $\dot{Q}_M$ and $\dot{Q}_R$ obviously depends not only on the energy spectrum (i.e., the relationship between $B_\mu$ and $J_{\mu\nu}$) but also on the temperature of the heat reservoir, which is particularly clear in Figs. \ref{Q_TM}(b), (c), (d), (f), and (h). $\dot{Q}_M$ also has a critical temperature $T_M^c$ that reverses the direction of the heat current. When the direction of the magnetic field changes, the critical temperature still exists, but the competition between $\dot{Q}_M$ and $\dot{Q}_R$ may disappear, as shown in Fig. \ref{Q_theta}.
\begin{figure}
	\centering
		\includegraphics[width=8.3cm]{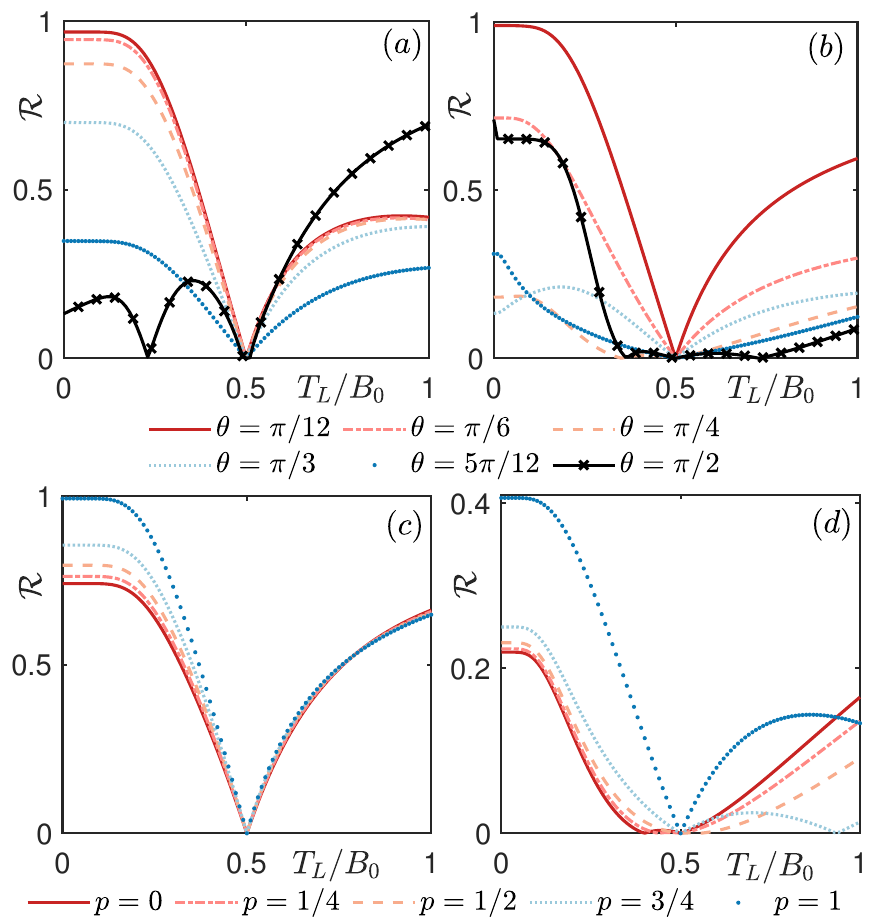}
	\caption{Rectification $\mathcal{R}$ with the temperature of the left heat reservoir $T_L$. (a) and (b) describe $\mathcal{R}$ versus the magnetic field direction $\theta$. (c) and (d) describe the dependence of $\mathcal{R}$ on the fraction $p$ when the magnetic field is the TF, i.e., the system's steady state depends on the initial state. The diode property described in the two figures on the left (or right) derives from the asymmetry of the free magnetic field $B_\mu$ (or of the coupling strength between the spins $J_{\mu\nu}$). Here, $B_0=1$, $\kappa_L=\kappa_R=0.001B_0$, $\kappa_M=0$, and $T_R=0.5B_0$. In (a) and (c), $B_L=3B_0$, $B_M=2B_0$, $B_R=B_0$, $J_{LM}=J_{MR}=0.1B_0$. In (b) and (d), $B_L=B_M=B_R=B_0$, $J_{LM}=B_0$, $J_{MR}=0.1B_0$.}
\label{R}
\end{figure}
\section{Quantum thermal diode}
\label{AppendixE}
An asymmetrical quantum system with two terminals can often be regarded as a quantum thermal diode, and the rectification $\mathcal{R}$ is used to measure the performance of the device. The rectification factor is defined as the ratio of the difference between the heat currents of the system before and after exchanging the temperatures of the two heat reservoirs to the maximum value between them, i.e.,
\begin{align}
\mathcal{R}_\mu=\frac{\vert\dot{Q}_\mu^f+\dot{Q}_\mu^r\vert}{\max[\dot{Q}_\mu^f,\dot{Q}_\mu^r]},\quad \mu=L,R,\label{Rdefinition}
\end{align}
where the superscript $f$ (or $r$) indicates the forward (or reverse) temperature bias. Due to $\dot{Q}_L=-\dot{Q}_R$ at the steady state, there is $\mathcal{R}_L=\mathcal{R}_R\equiv\mathcal{R}$.
According to Eq. (\ref{Rdefinition}), when $\mathcal{R}=0$, the system is perfectly symmetrical, i.e., there is no rectification effect; when $\mathcal{R}=1$, the system with the strongest asymmetry can be considered as a perfect quantum thermal diode; and when $0<\mathcal{R}<1$, the system shows specific rectification.

For the system where the middle spin is not in contact with the thermal reservoir, the variation of $\mathcal{R}$ with the temperature of the left thermal reservoir $T_L$ is given in Fig. \ref{R}.
Since the system is decoupled and both subsystems have vanished heat current at $\theta=s\pi$, this case will not be considered in this section. 
In Figs. \ref{R}(a) and (b), the dependences of the diode, which is based on the asymmetry of the magnetic field strength $B_\mu$ and the coupling strength between the spins $J_{\mu\nu}$, on the magnetic field direction $\theta$ are investigated respectively. We find that when the asymmetry of the system arises from the magnetic field of each spin $B_\mu$, the rectification decreases as the magnetic field direction changes from longitudinal to transverse as shown in Fig. \ref{R}(a). However, the variation of the rectification is no longer monotonic around $\theta=\frac{\pi}{2}$, and the heat current at $\theta=\frac{\pi}{2}$ is obtained by evolving the steady state of the nearest-neighbor point as the initial state. When the rectification stems from the coupling strength between the spins, we find that the dependence of $\mathcal{R}$ on $\theta$ is not monotonic, as depicted in Fig. \ref{R}(b).
Figures \ref{R}(c) and (d) depict the dependence of the diode on the fraction $p$ in the presence of the TF, in which the diode feature originated from $B_\mu$ and $J_{\mu\nu}$.
When the magnetic field of each spin is different, shown in Fig. \ref{R}(c), the system can still be regarded as a good diode at $p=1$. If only the coupling strengths between spins are different, shown in Fig. \ref{R}(d), the system has an imperfect rectification in the TF case.

\bibliography{quantum_thermal_device}

\end{document}